\documentclass[journal]{IEEEtran}

\ifCLASSINFOpdf

\else

\fi

\usepackage{color}

\usepackage{graphicx,tabularx,array,amsmath,amsthm,thmtools,subcaption}
\usepackage{enumitem}
\usepackage{mathtools}

\usepackage{caption}
\usepackage{amsfonts}
\usepackage{bm}
\usepackage{bbm}
\usepackage{makecell}
\usepackage{multirow}
 \usepackage{amssymb}
\usepackage{amsthm}
\usepackage{txfonts}
\usepackage[T1]{fontenc}
\usepackage[scr=dutchcal]{mathalfa}
\let\mathscr\mathbscr

\usepackage{cite}
\usepackage{amsmath}
\usepackage{booktabs}
\usepackage{cuted}
\usepackage{algorithm}
\usepackage{algorithmic}
\allowdisplaybreaks % makes automatic splits of equations

\declaretheoremstyle[headfont=\bfseries, 
    bodyfont=\normalfont]{normalhead}
\declaretheorem[style=normalhead]{Example}

\newtheorem{Theorem}{Theorem}

\newtheorem{Lemma}{Lemma}
\newtheorem{Corollary}{Corollary}
\newtheorem{Remark}{Remark}
\newtheorem{Definition}{Definition}

\DeclareMathOperator*{\argmax}{argmax}

\begin{document}

%\newgeometry{top=1in,bottom=0.75in,right=0.75in,left=0.75in}
%---------------------------------------------

% \begin{titlepage}
% \end{titlepage}

\title{A General Framework for Temporal Fair User Scheduling in NOMA Systems}

%\title{Opportunistic Multi-User Scheduling under Temporal Fairness Constraints (working title)}

% \title{Opportunistic Multiuser Scheduling with Temporal Fairness}

% author names and affiliations
% use a multiple column layout for up to three different
% affiliations

\author{
\IEEEauthorblockN{Shahram Shahsavari, Farhad Shirani, Elza Erkip\\}
\IEEEauthorblockA{Department of Electrical and Computer Engineering\\
New York University
\\
\{shahram.shahsavari,fsc265,elza\}@nyu.edu}
}
\maketitle

% As a general rule, do not put math, special symbols or citations
% in the abstract
\begin{abstract}
\let\thefootnote\relax\footnotetext{This work is supported by NYU WIRELESS Industrial Affiliates and National Science Foundation grants EARS-1547332 and NeTS-1527750.}
Non-orthogonal multiple access (NOMA) is one of the promising radio access
techniques for next generation wireless networks. Opportunistic multi-user scheduling is necessary to fully exploit multiplexing gains in NOMA systems, but compared with traditional scheduling, inter-relations between users' throughputs induced by multi-user interference poses new challenges in the design of NOMA schedulers.  A successful NOMA scheduler has to carefully balance the following three objectives: maximizing average system utility, satisfying desired fairness constraints among the users and enabling real-time, and low computational cost implementations. 
%First, the scheduler should be implementable in real-time with reasonable computational cost. Second, resource allocation between the users should satisfy the desired fairness constraints. Third, the average system utility needs to be maximized.  
 In this paper, scheduling for NOMA systems under temporal fairness constraints is considered. Temporal fair scheduling leads to communication systems with predictable latency as opposed to utilitarian fair schedulers for which latency can be highly variable. 
It is shown that optimal system utility is achieved using a class of opportunistic scheduling schemes called \textit{threshold based strategies} (TBS). %Additionally, it is shown that any optimal scheduling strategy for NOMA systems can be written in the form of a TBS. 
One of the challenges in temporal fair scheduling for heterogeneous NOMA scenarios - where only specific users may be activated simultaneously - is to determine the set of feasible temporal shares. In this work, a variable elimination algorithm is proposed to accomplish this task. 
Furthermore, an (online) iterative algorithm based on the Robbins-Monro method is proposed to construct a TBS by finding the optimal thresholds for a given system utility metric. %A novel perturbation technique is proposed which circumvents the optimization of the tie-breaking decision rule used in TBS. The method provides TBSs whose average system utility is arbitrarily close to the utility achieved using optimal tie-breaking decision rules and thresholds. 
Various numerical simulations of practical scenarios are provided to illustrate the effectiveness of the proposed NOMA scheduling in static and mobile scenarios.
\end{abstract}

% no keywords

\section{Introduction}

Non-orthogonal multiple access (NOMA) has emerged as one of the key enabling technologies for fifth generation wireless networks \cite{DaiSurvey2018,saito2013system,ding2017application,ding2017survey}. In order to satisfy the ever-growing demand for higher data rates in modern cellular systems, NOMA proposes serving multiple users in the same resource block. This is in contrast with conventional cellular systems which operate based on orthogonal multiple access (OMA) techniques such as orthogonal frequency-devision multiple access (OFDMA) \cite{seong2006optimal}. In OMA systems, each time-frequency resource block is assigned to only one user in each cell, whereas, in NOMA systems, multiple users can be scheduled either in uplink (UL) or in downlink (DL) simultaneously \cite{ding2016general}. As a result, the scheduler in the NOMA system may choose among a larger collection of users at each resource block as compared to an OMA scheduler, often leading to a higher system throughput \cite{otao2012performance}. The high system  throughput is due to NOMA multiplexing gains, achieved through a combination of  superposition encoding strategies at  the transmitter(s) and successive interference cancellation (SIC) decoding at the receiver(s) \cite{otao2012performance, Islam2018,yu2004sum}. 
 However, the inter-relations between users' throughputs induced by multi-user interference complicate the design of high-performance schedulers, giving rise to new challenges both in terms of designing user power allocation schemes \cite{lei2016power,liu2016fairness,yang2016general} as well as optimal schedulers \cite{saito2013system,cui2018optimal}.
 %{\color{red}(remove this sentence? makes it look easy) As an example, a well-designed NOMA  scheduler for a single-cell downlink scenario pairs a user with a weak channel condition with a user with a strong channel condition \cite{liang2017user,ding2016impact}.} 
 %The additional degree of freedom of NOMA due to activating multiple users 
  %There is a large body of work on the design of multi-terminal communication strategies which maximize the aforementioned multiplexing gains.
Ideally, the scheduler is designed in tandem with the encoding and decoding strategies and power optimization techniques. However, due to the complexity of the problem, scheduling is usually studied in isolation assuming that the system throughputs are given to the scheduler based on a predetermined communication strategy \cite{saito2013non,saito2013system}. 

The objective of a NOMA scheduler is to maximize the system utility (e.g. system throughput) subject to the users' individual demand constraints, e.g. temporal demands or minimum utility demands. More precisely, at each resource block, the scheduler estimates the set of resulting system utilities from activating any specific subset of uplink or downlink users. It then  chooses the set of active users in that block based on this information and users' individual fairness demands. 
% \begin{figure}[!t]
%  \centering \includegraphics[width=\linewidth, draft=false]{Plot4}
%  \caption{The red(dashed blue) graph corresponds to the resulting throughput from activating user one(two) in an OMA communication setup. A and B show the order in which users are activated over consecutive time-frequency blocks by a utilitarian fair scheduler and a temporally fair scheduler, respectively. Each red(blue stripes) block corresponds to activating user one(two) in that resource block.  
%  %
%  %
%  }
%  \label{Fig:Plot2}
% \end{figure} 
Quantifying fairness in user scheduling has been a topic of significant interest. Various criteria on the users' quality of service (QoS) have been proposed to model and evaluate fairness of scheduling strategies. For OMA systems, scheduling under utilitarian \cite{liu-elsevier,zhang2008opportunistic}, proportional \cite{kelly1998rate,viswanath2002opportunistic}, and temporal  \cite{liu-jsac,kulkarni2003opportunistic} fairness criteria have been studied.

In delay sensitive applications, a system with reasonable and predictable latency may be more desirable than a system with highly variable latency, but potentially higher throughput. In such scenarios, temporally fair schedulers are often favored over utilitarian fair schedulers. 
Temporally fair schedulers provide each user with a minimum temporal share in order to control the average delay \cite{issariyakul2004throughput}. 
Furthermore, most of the power consumption in cellular devices is due to the radio electronics which are activated during data transmission and reception. Consequently, the maximum power drain of users can be restricted by considering upper-bounds on the users' temporal shares \cite{kulkarni2003opportunistic}. From the perspective of the network provider, an additional upside of temporally fair schedulers is that users with low channel quality do not hinder network throughput as severely as in utilitarian fair schedulers \cite{asadi2013survey}.
% From the perspective of the network provider, an additional upside of temporally fair schedulers is that users with low channel quality do not hinder network throughput as severely as in utilitarian fair schedulers \cite{asadi2013survey}.
%  Moreover, it is relatively easier to determine whether a scheduling scheme satisfies temporal fairness by means of monitoring time-use of users compared to other fairness criteria such as proportional fairness \cite{shahram-journal}. 
 There has been a significant body of work dedicated to the study of temporally fair schedulers in wireless local area networks (WLAN)\cite{tan2004time,joshi2008airtime} and OMA cellular systems \cite{shahram-letter,kulkarni2003opportunistic,shahsavari2018joint}. However, temporal fairness of NOMA schedulers, which is the topic of this paper, has not been investigated before.

%Prior works have considered temporal fair scheduling for OMA systems \cite{asadi2013survey,liu-jsac,liu-elsevier}. 
%

In OMA systems, optimal utility subject to temporal demand constraints is achieved using a class of opportunistic scheduling strategies called \textit{threshold based strategies} (TBS) \cite{liu-jsac}.
An opportunistic scheduler exploits the time-varying nature of the users' wireless channels. In TBSs, at each resource block, the active user $u_i$ is chosen based on the sum of two components: i) Performance value $R_i$ (tipically transmission rate), and ii) a constant term called the \textit{user threshold} $\lambda_i$. The user thresholds are chosen to optimize the tradeoff between system utility and users' temporal share demands. The thresholds can be interpreted as the Lagrangian multipliers corresponding to the fairness constraints in the optimization of the system utility. 
In \cite{liu-jsac}, a method based on the Robbins-Monro algorithm is proposed to construct optimal temporally fair TBSs for OMA systems.
%{\color{red} I shortened the next blue part and moved it to the next parag. pls check.} {\color{blue} The question of existence and construction of optimal NOMA TBSs is more challenging. A NOMA TBS assigns a threshold to each subset of users which can be activated simultaneously. In an optimal TBS the thresholds assigned to the subsets of users are inter-related. As an example, we show that in the two user NOMA system an optimality achieving TBS assigns a threshold to the joint user $\{u_1,u_2\}$ which is equal to  the sum of the thresholds of the two individual users $u_1$ and $u_2$. 
%Consequently, the existence of an optimal TBS is contingent upon the existence of a set of Lagrangian multipliers (thresholds) satisfying the optimality conditions in maximizing the system utility given the inter-relations mentioned above.}  

In this work, we consider the user scheduling problem for NOMA systems under temporal fairness constraints. 
%Each user is required to be activated for at least a predetermined fraction of resource blocks.
We provide a mathematical formulation of the problem which is applicable under general utility models and assumptions on the subsets of users which can be activated simultaneously. Our model is applicable to both UL and DL scenarios. 

We first address the question of feasibility of a set of temporal demands in a given NOMA system. A vector of temporal shares is said to be \textit{feasible} if there exists a scheduling strategy for which the resulting user temporal shares are equal to the elements of the vector. In OMA systems, since exactly one user is active at each block, a vector of temporal shares is feasible as long as the its elements sum to one.  However, in NOMA systems the set of feasible temporal shares is not trivially known. Determining the feasible set is especially challenging in large heterogeneous NOMA systems, where only specific users may be activated simultaneously. In Section \ref{sec:exist}, we propose a variable elimination method to derive the set of feasible temporal shares in arbitrary heterogeneous NOMA scenarios. Furthermore, we prove that given a feasible set of temporal demands, TBSs are optimal for NOMA systems. We further prove that any optimal scheduling strategy can be written in the form of a TBS. 

The question of existence and construction of optimal NOMA TBSs is more challenging than OMA TBSs. The reason is that a NOMA TBS assigns a threshold to each subset of users which can be activated simultaneously, rather than each user separately. Therefore, in an optimal TBS the thresholds assigned to the subsets of users are inter-related. In Section \ref{sec:robbins-monro}, we propose a construction method based on the Robbins-Monro algorithm to find the optimal thresholds for a NOMA TBS.
In Section \ref{sec:discrete},  we consider practical NOMA systems where discrete modulation and coding strategies are used. In this case, the resulting system utilities are staircase functions of the users' signal to noise ratios. As a result, the utility from activating different subsets of users may lead to a tie in the TBS decision. This necessitates the design and optimization of a tie-breaking decision rule \cite{liu-jsac}. We propose a perturbation technique which circumvents the optimization and leads to TBSs whose average system utility is arbitrarily close to the optimal utility. In Section \ref{sec:sim}, we provide simulations and numerical examples in several practical scenarios involving static and mobile settings. We observe that the proposed scheduling algorithm adapts to the changes due to user mobility under typical velocity assumptions.

\section{Notation}
 We represent random variables by capital letters such as $X, U$. Sets are denoted by calligraphic letters such as $\mathcal{X}, \mathcal{U}$. The set of natural numbers, and the real numbers are shown by $\mathbb{N}$, and $\mathbb{R}$ respectively.   The set of numbers $\{1,2,\cdots, n\}, n\in \mathbb{N}$ is represented by $[n]$. 
 The vector $(x_1,x_2,\cdots, x_n)$ is written as $x^n$. The $m\times t$ matrix $[g_{i,j}]_{i\in [m], j\in [t]}$ is denoted by $g^{m\times t}$.
 For a random variable $X$, the corresponding probability space is $(\mathcal{X}, \mathbf{F}_{X}, P_X)$, where $\mathbf{F}_X$ is the underlying $\sigma$-field. The set of all subsets of $\mathcal{X}$ is written as $2^{\mathcal{X}}$. For an event $\mathcal{A}\in 2^{\mathcal{X}}$, the random variable $\mathbbm{1}_{\mathcal{A}}$ is the indicator function of the event.  We write $X\sim Unif[a,b]$ for a random variable $X$ uniformly distributed on the interval $[a,b]$.
 Families of sets are shown using sans-serif letters (e.g. $\mathsf{X}=2^{\mathcal{X}})$. The closed interval $\{x: a\leq x\leq b\}$ is shown by $[a,b]$. %The notation $\floor{x}$ ($\ceil{x}$) is used to represent the closest integer smaller (bigger) than $x$. 
 Finally, $mod_k(i), i,k\in \mathbb{N}$ represents the value of $i$ modulo $k$.

% \section{Problem Formulation}

% In this section, we formulate multi-user scheduling under temporal fairness constraints. We consider a single cell time-slotted system with multiple users distributed within the cell. It is assumed that the users have individual temporal demands, e.g. it is required that each user be activated for at least  a predefined fraction of the time-slots. At each time slot, the base station activates a subset of the users simultaneously. The set of all permissible subsets of users - also called the set of \textit{virtual users} - depends on the system under consideration. For example, in full-duplex (FD) systems, a pair consisting of an uplink and  a downlink user can be activated at each time slot; whereas in wireless communication using NOMA, the scheduler may activate multiple downlink or multiple uplink users simultaneously. The system utility due to activating any specific virtual user in a given time-slot is called the performance value of that virtual user in that time-slot. As an example, if throughput is considered as the system utility, then the performance value of each virtual user is a function of the channel gains of the users as well as the decoding scheme (e.g. SIC). The (temporal fair) multi-user scheduling problem is formalized as follows. 
{
\section{System Model} \label{sec:sys-model}
In this section, we describe the system model and formulate NOMA multi-user scheduling under temporal fairness constraints. We consider a single-cell time-slotted system with $n$ users distributed within the cell. We define the user set as $\mathcal{U}=\{u_1,u_2,\cdots,u_n\}$ where $u_i, i\in[n]$ denotes the $i$th user. 
At each time-slot, the base station (BS) activates a subset of { UL or DL} users simultaneously using NOMA. The maximum number of active users at each time-slot is bounded from above due to practical considerations such as latency and computational complexity at the decoder. For example, the decoding complexity and communication delay under SIC is proportional to the number of multiplexed users \cite{DaiSurvey2018}. Consequently, only subsets of users with at most $N_{max}\leq n$ elements can be activated simultaneously, where $N_{max}$ is determined based on the communication setup under consideration. Several works on NOMA scheduling consider $N_{max}=2$ and $N_{max}=3$ under various utilitarian and proportional fairness constraints \cite{wei2016power,liang2017user}. 
A subset of users which can be activated simultaneously is called  a \textit{virtual user}.

\begin{Definition}[\bf{Virtual User}] \label{def:virtual-user}
For a NOMA system with $n$ users and maximum number of active users $N_{max}\leq n$, the set of virtual users is defined as
\[\mathsf{V}=\left \{\mathcal{V}_j\big|j\in [m]\right\}= \left\{\mathcal{V}_j\subset \mathcal{U}\big| |\mathcal{V}_j|\leq N_{max}\right\}. \]
The set $\mathcal{V}_j, j\in [m]$ is called the $j$th virtual user.  The total number of virtual users in a NOMA system is equal to $m = {n \choose 1}+{ n \choose 2}+\cdots+{ n \choose N_{max}}$.
\end{Definition}

%The set of all permissible subsets of users, virtual users, depends on the system under consideration. For example, in full-duplex (FD) systems, a pair consisting of an uplink and a downlink user can be activated at each time slot; whereas in wireless communication using NOMA, the scheduler may activate multiple downlink or multiple uplink users simultaneously. we assume that $|\mathcal{V}_j|\leq n_{max}, j\in[m]$, i.e. only virtual users consisting of $n_{max}$ or fewer users are permissible. 
 %Although we focus on the NOMA downlink scenario in this paper, the same framework can be applied to NOMA uplink as well as FD networks.
%The following example explains Definition \ref{def:virtual-user} by considering a three-user scenario where $n_{max}=2$.
%{\color{red} do we need this example? it seems trivial}
%\begin{Example}
%Consider a three user downlink scenario, i.e. $n=3$. The user set is $\mathcal{U}=\{u_1,u_2,u_3\}$. Let $N_{max}=2$, that is, the BS can activate any subset of two or less users  at any given time. There are six virtual users: $\mathcal{V}_1=\{u_1\},\mathcal{V}_2=\{u_2\},\mathcal{V}_3=\{u_3\},\mathcal{V}_4=\{u_1,u_2\},\mathcal{V}_5=\{u_1,u_3\},\mathcal{V}_6=\{u_2,u_3\}$. For example, activating virtual user $\mathcal{V}_3$ in a given time-slot is equivalent to activating $u_1$ and $u_2$ in that time-slot. 
%\label{Ex:1}
%\end{Example}

Our objective is to design a scheduler which maximizes the average network utility subject to  temporal fairness constraints. At the beginning of each time-slot, the scheduler finds the utility due to activating each of the virtual users, and decides which virtual user to activate in that time-slot. The utility is usually defined as a function of the throughput of the elements of the virtual user. 
%The resulting utility due to activating virtual user $\mathcal{V}_j, j\in [m]$ at time-slot $t$ is denoted by $R_{j,t}$.  The vector $(R_{1,t},R_{2,t}, \cdots, R_{m,t}), t\in \mathbb{N}$ is called the performance vector at time $t$. 

\begin{Definition}[\bf{Performance Vector}]
The vector of jointly continuous variables $(R_{1,t},R_{2,t}, \cdots, R_{m,t}), t\in \mathbb{N}$ is the performance vector of the virtual users at time $t$. The sequence $(R_{1,t},R_{2,t}, \cdots, R_{m,t})$ is a sequence of independent\footnote{Note that the realization of the performance vector is independent over time, however the performance of the virtual users at a given time-slot may be dependent on each other.} vectors distributed identically according to the joint density $f_{R^m}$.
\end{Definition}
\begin{Remark}
It is assumed that the performance vector is bounded with probability one. Alternatively, we assume that $P(R^m\in [-M,M]^m)=1$ for large enough $M\in \mathbb{R}_{\geq 0}$.
\label{Rem:bound}
\end{Remark}
\begin{Remark}
 For the virtual user $\mathcal{V}_j= \{u_{i_1},u_{i_2},\dots, u_{i_{k_j}}\}, j\in [m], k_j\in [n], i_1,i_2,\cdots,i_{k_j}\in [n]$, we sometimes write $\mathcal{V}_{i_1,i_2,\cdots, i_{k_j}}$ ($R_{i_1,i_2,\cdots, i_{k_j}}$) instead of $\mathcal{V}_j$ ($R_j$) to represent the virtual user (performance variable).
\end{Remark}
The following example clarifies the notion of performance vector and provides a characterization for $(R_{1,t},R_{2,t}, $ $\cdots, R_{m,t}), t\in \mathbb{N}$ in a large class of practical applications. 

\begin{Example}
 In this example, we explicitly characterize the performance vector of a NOMA downlink system at any time-slot, where the system utility is defined to be the transmission sum-rate. The characterization can also be used for NOMA uplink scenarios with minor modifications. Let $h_{i,t}$ be the propagation channel coefficient between user $u_i$ and the BS which captures small-scale and large-scale fading effects \cite{rappaport}.
%  We use the standard channel model , where at time $t$, the input $[X_{i,t}]_{i\in [n]}$ and the output $[Y_{i,t}]_{i\in [n]}$ are related through the following additive Gaussian channel $Y_{i,t}=h_{i,t}\sum_{k\in [n]}X_{k,t}+N_{i,t}, i\in [n]$. It assumed that $[N_{i,t}]_{i\in [n]}$ is a vector of Gaussian noise variables each  of which have variance $\sigma^2$. 
 It is also assumed that the channel coefficients $h_{i,t}, i\in [n]$ are independent over time. 
%The BS can estimate the channel coefficients  $h_{i,t}, i\in [n]$  between the BS and user $u_i$ at time-slot $t$.
%, where $h_{i,t}=\sqrt{\beta_{i,t}}g_{i,t}\in \mathbb{C}$, and $\beta_{i,t}\in \mathbb{R^+}$ and $g_{i,t}\in \mathbb{C}$ are the large-scale and the small-scale channel coefficients, respectively. Under the Rayleigh block fading model $g_{i,t}, t\in \mathbb{N}$ is a sequence of independent and identically distributed complex Gaussian random variables with unit variance. The parameter $\beta_{i,t}$ models range dependent path-loss and shadow fading and is assumed to be constant over the large-scale coherence time of the channel. 
 Let $R_{j,t}, j\in [n], t\in \mathbb{N}$ be the sum-rate of the elements of virtual user $\mathcal{V}_j$ given that it is activated at time $t$. In NOMA downlink, a combination of superposition coding at the BS and SIC decoding at the user equipment has been proposed \cite{islam2017power}. As envisioned for practical NOMA downlink systems, the decoding occurs in the order of increasing channel gains \cite{otao2012performance}. For a fixed virtual user $\mathcal{V}_j, j\in [m]$ and user $u_i\in \mathcal{V}_j$, let  $\mathcal{I}_{i,j,t}$  be the set of elements of $\mathcal{V}_j$ whose channels are stronger than that of $u_i$ at time $t$. Alternatively, define  $\mathcal{I}_{i,j,t}=\{u_l \in \mathcal{V}_j\big| |h_{l,t}|>|h_{i,t}|\}$. If $\mathcal{V}_j$ is activated at time $t$, user $u_i$ applies SIC to cancel the interference from users in $\mathcal{V}_j$ whose channel gain is lower than that of $u_i$, hence only the signals from users in $\mathcal{I}_{i,j,t}$ are treated as noise and result in a lower transmission rate for $u_i$. It is well-known that this decoding strategy is optimal when the users' channels are degraded \cite{ElGamalLec}. 
 The interested reader is referred to \cite{yu2004sum} for a detailed description of optimal decoding strategies in the downlink scenario.  
 The signal to interference plus noise ratio (SINR) of user $u_i$ is 
\begin{align}
\text{SINR}_{i,j,t}=\frac{P_{i,j,t}|h_{i,t}|^2}{\sum_{l\in \mathcal{I}_{i,j,t}}P_{l,j,t}|h_{l,t}|^2+\sigma^2}, i \in \mathcal{V}_j, j\in[m], t\in \mathbb{N}, \label{eq:sinr}
\end{align}
where, $P_{i,j,t}$ denotes the transmit power assigned to $u_i$ if virtual user $\mathcal{V}_j$ is activated at time-slot $t$, and $\sigma^2$ is the noise power.
Let $R_{i,j,t}$ denote the rate of user $u_i$ if virtual user $\mathcal{V}_j$ is activated at time-slot $t$, and let $R_{j,t}$ be the resulting sum-rate. Then,
\begin{align}
&R_{i,j,t}=\log_2\left(1+\text{SINR}_{i,j,t}\right), i \in \mathcal{V}_j, j\in[m], \label{eq:user-rate}\\
&R_{j,t}=\sum_{i=1}^n R_{i,j,t} \mathbbm{1}_{\left\{u_i \in \mathcal{V}_j\right\}},j\in[m]. \label{eq:vitual-rate}
\end{align}
\label{Ex:NOMA}
\end{Example}
% The system utility due to activating any specific virtual user in a given time-slot is called the performance value of that virtual user in that time-slot. As an example, if throughput is considered as the system utility, then the performance value of each virtual user is a function of the channel gains of the users as well as the decoding scheme (e.g. SIC). The (temporal fair) multi-user scheduling problem is formalized as follows. 
Temporal fair scheduling guarantees that each user is activated for at least a predefined fraction of the time-slots. More precisely, user $u_i, i\in [n]$ is activated for at least $\underline{w}_i$ of the time, where $\underline{w}_i\in [0,1]$. 
%The vector $\underline{w}^n$ is called the vector of lower temporal share demands of the users. 
Similarly, the scheduler guarantees that the users are not activated more than a predefined fraction of the time-slots which are given by the upper temporal share demands $\overline{w}^n$.

\begin{Definition}[\bf{Temporal Demand Vector}]
For an $n$ user NOMA system, the vector $\underline{w}^n$ ($\overline{w}^n)$ is called the lower (upper) temporal demand vector. 
\end{Definition}

At time-slot $t\in \mathbb{N}$, the scheduling strategy takes  $(R_{1,k},R_{2,k}, \cdots, R_{m,k}), k\in [t]$, the independent realizations of the performance vector in all time-slots up to time $t$ and outputs the virtual user which is to be activated in the next time-slot. The NOMA scheduling setup is parametrized by 
$(n,N_{max}, \underline{w}^n, \overline{w}^n, f_{R^m})$.

\begin{figure*}
\centering 
\includegraphics[width=.55\textwidth, draft=false]{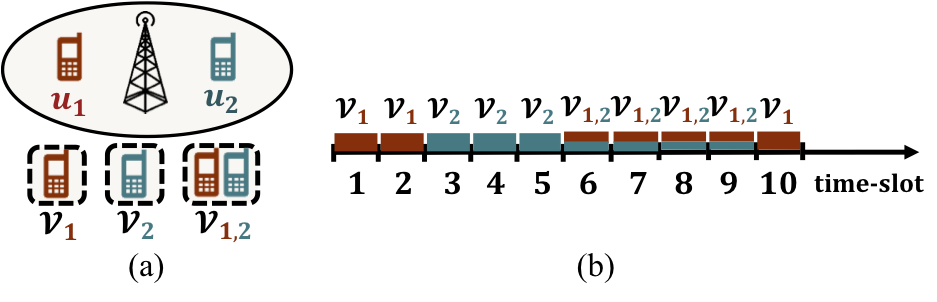}
\caption{(a) Two-user NOMA system with three virtual users, (b) Weighted Round Robin scheduling strategy when $\underline{w}_1=\underline{w}_2=0.6$.}
\label{Fig:NOMA2}
\end{figure*}

%The objective in the study of the scheduling problem is to determine which virtual user is activated at each time slot so as to maximize average network utility subject to fairness constraints. The scheduling strategy $Q_t$ determines the virtual user which is activated at time $t$ as a function of the independent realizations of the performance vector $R^m$ in all of the previous time slots. 

% The objective in the scheduling problem is to determine which virtual user is activated at each time slot subject to fairness and utility constraints. The scheduling strategy $Q$ determines the virtual user which is activated at time $t$ as a function of the independent realizations of the performance vector $R^m$ in all of the previous time slots. 

%More precisely, let $R^{m\times t}$ be the matrix of performance vectors up until time $t$, where the performance vectors at different time slots are independent and identically distributed. Then, $Q_t(R^{m\times t})$ is the virtual user activated at that time slot.  

%At time $t$ a generic scheduling strategy Q gets all the performance vectors until time  selects a virtual    A strategy for the scheduling problem is defined below. 
\begin{Definition}[\bf{Scheduling Strategy}]
A scheduling strategy (scheduler) $Q= (Q_t)_{t\in \mathbb{N}}$ for the scheduling setup parametrized by the tuple $(n,N_{max}, \underline{w}^n, \overline{w}^n, f_{R^m})$ is a family of (possibly stochastic) functions $Q_t: \mathbb{R}^{m\times t} \to \mathcal{V}, t\in \mathbb{N}$, for which:
\begin{itemize}
    \item { The input to $Q_t, t\in \mathbb{N}$ is the
matrix of performance vectors $R^{m\times t}$ which consists of $t$ independently and identically distributed column vectors with distribution} $f_{R^m}$.% The vector $(R_{1,k}, R_{2,k},\cdots, R_{m,k}), k\in [t]$ is an independent realization of the performance vector $R^m$ corresponding to time-slot $k$. 
\item{ The temporal demand constraints are satisfied:
\begin{align}
P\left(\underline{w}_i-\epsilon\leq \underline{A}_i^{Q} \leq \overline{w}_i+\epsilon, i\in [n]\right)=1, \forall \epsilon>0,
\label{Def:tem_fair}
\end{align}}
\end{itemize}
where, the temporal share of user $u_i, i\in [n]$ up to time $t\in \mathbb{N}$ is defined as
\begin{align}
A^Q_{i,t}=\frac{1}{t}\sum^t_{k=1}\mathbbm{1}_{\big\{u_i\in Q_k(R^{m\times k})\big\}}, \forall i\in [n], t\in \mathbb{N},
\label{Def:temp_share}
\end{align}
and the average temporal share of user $u_i, i\in [n]$ is
\begin{align}
&\underline{A}_i^{Q}=\liminf_{t\rightarrow \infty} A_{i,t}^{Q}, \forall i\in [n].
\label{Def:av_temp_share}
\end{align}

\end{Definition}

\begin{Remark}
Note that analogous to Equation \eqref{Def:av_temp_share}, one could define $\overline{A}_i^{Q}=\limsup_{t\rightarrow \infty} \overline{A}_{i,t}^{Q}, \forall i\in [n]$ and modify Equation \eqref{Def:tem_fair} accordingly. However, as we will show in the next sections, for scheduling strategies of interest, we have $\underline{A}_i^{Q}=\overline{A}_i^{Q}$.
\end{Remark}
\begin{Remark}
A scheduling setup where the temporal shares of users are required to take a specific value, i.e. $\underline{A}_i^Q=w_i, i\in [n]$, is called a scheduling setup with equality temporal constraints and is parametrized by $(n,N_{max}, {w}^n, {w}^n, f_{R^m})$.
\end{Remark}

\begin{Definition}[\bf{System Utility}]
For a scheduling strategy $Q$:
\begin{itemize}
    \item { The average system utility up to time t, is defined as 
\begin{align}
U^Q_t&=\frac{1}{t}\sum^t_{k=1}\sum_{j=1}^m R_{j,k}\mathbbm{1}_{\big\{Q_k(R^{m\times k})=\mathcal{V}_{j}\big\}}.
\label{Def:sys_utility}
\end{align}}
\item{The average system utility is defined as
\begin{align}
U^{Q}= \liminf_{t\rightarrow \infty} U^Q_t.
\label{Def:av_sys_utility}
\end{align}}
\end{itemize}

\end{Definition}
The strategy $Q=(Q_t)_{t\in \mathbb{N}}$ takes the matrix $R^{m\times t}$ of performance values up to time $t$ as input and outputs the virtual user $\mathcal{V}_j, j\in [m]$ which is to be activated at time $t$. Temporal share $A_{i,t}^{Q}, i\in [n],t\in \mathbb{N}$ in \eqref{Def:temp_share} represents the fraction of time-slots in which user $u_i$ is activated until time $t$. The variable $\underline{A}^Q_i, i\in [n]$ is an asymptotic lower bound to the temporal share of user $u_i$ and \eqref{Def:tem_fair} represents the temporal fairness constraints. Furthermore, $U^Q_t$ and $U^{Q}$ are the instantaneous and average system utilities, respectively. The objective is to design a scheduling strategy which achieves the maximum average network utility while satisfying temporal fairness constraints.

\begin{Remark}
\label{Rem:perf}
Typically, it is assumed that the scheduler does not know the distribution function $f_{R^m}$ prior to the start of scheduling. Rather, at each time-slot $t$, the scheduler uses its current and prior empirical observations of realizations of $R^m$ to choose the active virtual user.  For instance, the scheduler in Example \ref{Ex:NOMA} estimates the users' channels at the start of each time-slot and uses the estimate to choose the active user.  
\end{Remark}

%The strategy $Q=(Q_t)_{t\in \mathbb{N}}$ takes the matrix $R^{m\times t}$ of performance values until time $t$ as input and outputs the virtual user $\mathcal{V}_j, j\in [m]$ which is to be activated at time $t$. Temporal share $A^{i,t}_{Q}, i,t\in \mathbb{N}$ in \eqref{Def:temp_share} represents the fraction of time-slots in which user $u_i$ is activated until time-slot $t$. The variable $\underline{A}_Q^i, i\in [n]$ is an asymptotic lower bound to the temporal share of user $u_i$ and \eqref{Def:tem_fair} represents the temporal fairness constraints. Furthermore, $U_{Q,t}$ and $U_{Q}$ are the instantaneous and average system utilities, respectively. The objective is to design a scheduling strategy which achieves the maximum average network utility while satisfying temporal fairness constraints. 
\begin{Definition}[\bf{Optimal Strategy}]
For the scheduling setup parametrized by $(n,N_{max}, \underline{w}^n, \overline{w}^n, f_{R^m})$, a strategy $Q^*$ is optimal if and only if 
\begin{align}
Q^*\in\argmax_{Q\in \mathcal{Q}} U_{Q},
\label{Def:optimal}
\end{align}
where $\mathcal{Q}$ is the set of all strategies for the scheduling setup.
\end{Definition}

% \begin{Remark}
% Note that in general, an optimal scheduling strategy might not exist %since the space $\mathcal{Q}$ is not a compact metric space. 
% since the maximum in Equation \eqref{Def:optimal} may not exist. Furthermore, for some values of $(\underline{w}^n, \overline{w}^n)$, the set $\mathcal{Q}$ is empty.  However, for the formulation considered in this work, we will show that such an optimal strategy always exists given that $\mathcal{Q}$ is not empty. However, the optimal strategy may not be unique. 
% \label{Rem:fes}
% \end{Remark}

\begin{Example}
Consider the downlink scenario shown in Figure \ref{Fig:NOMA2}. In this scenario $n=2$, and $\mathcal{U}=\{u
_1,u_2\}$. Furthermore, $m=3$ and $\mathsf{V}=\{\mathcal{V}_1,\mathcal{V}_2,\mathcal{V}_3\}$, where $\mathcal{V}_1=\{u_1\},\mathcal{V}_2=\{u_2\}$, and $\mathcal{V}_3=\mathcal{V}_{1,2}=\{u_1,u_2\}$.
Let the fairness constraints be given by the temporal weight demands $\underline{w}_1=\underline{w}_2=0.6$ and $\overline{w}_1=\overline{w}_2=1$. This requires each user to be activated for at least 0.6 fraction of the time.  One possible scheduling strategy in this scenario is the Weighted Round Robin (WRR) strategy shown in Figure \ref{Fig:NOMA2}(b). The strategy is described below:
\begin{align}\label{eq:round-robin}
Q_t=
\begin{cases}
\mathcal{V}_{1}, &\mbox{if ~} 0\leq mod_{10}(t)\leq 2,\\
\mathcal{V}_{2}, &\mbox{if ~} 3\leq mod_{10}(t)\leq 5,\\
\mathcal{V}_{1,2}, &\mbox{if ~} 6\leq mod_{10}(t)\leq 9.
\end{cases}
\end{align}

 The WRR strategy is a non-opportunistic strategy where virtual users are chosen independently of the realization of the performance vector. As a result, the temporal share $A_{i,t}^Q, i\in [n], t\in \mathbb{N}$ is a deterministic function of $t$. Note that $\underline{A}_i^{Q}=0.3+0.4=0.7, i=1,2$; hence the WRR strategy satisfies the temporal demand conditions \eqref{Def:tem_fair}. Also, it is straightforward to show that the average network utility is $U^Q=0.3\mathbb{E}(R_1)+0.3\mathbb{E}(R_2)+0.4\mathbb{E}(R_3)$. 
%\textcolor{red}{Shahram:} Please add Round Robin as example to explain the above definition and notation.
\label{Ex:rep}
\end{Example}

%\begin{Example}
%Consider the NOMA system described in example 1. In this example, we focus on a Weighted Round Robin (WRR) scheduling strategy which is non-opportunistic. Since the minimum temporal weight is $0.6$ for each user, virtual user $\mathcal{V}_{3}$ has to be chosen at least for $\underline{w}_1+\underline{w}_2-1=0.2$ fraction of time-slots. Let 
%\begin{align}\label{eq:round-robin}
%Q_t=
%\begin{cases}
%\mathcal{V}_{1}, &\mbox{if ~} 0\leq mod(t,10)\leq 2,\\
%\mathcal{V}_{2}, &\mbox{if ~} 3\leq mod(t,10)\leq 5,\\
%\mathcal{V}_{3}, &\mbox{if ~} 6\leq mod(t,10)\leq 9.
%\end{cases}
%\end{align}
%Although $A^{i,t}_Q$ provided in \eqref{Def:temp_share} is potentially a random variable given $i$ and $t$, it is a deterministic value in this example, since the strategy is deterministic and independent of performance values. Furthermore, it can be shown that $\underline{A}^i_{Q}=0.3+0.4=0.7, i=1,2$; hence the WRR strategy $Q$ provided in \eqref{eq:round-robin} satisfy temporal demand conditions \eqref{Def:tem_fair}. Also, applying Law of Large Numbers (LLN) to \eqref{Def:sys_utility}, it can be shown that the average network utility introduced in \eqref{Def:av_sys_utility} is $U_Q=0.3\mathbb{E}[R_1]+0.3\mathbb{E}[R_2]+0.4\mathbb{E}[R_3]$. 
%\textcolor{red}{Shahram:} Please add Round Robin as example to explain the above definition and notation.
%\end{Example}

 The set $\mathcal{Q}$  includes strategies with memory as well as non-stationary and stochastic strategies. As a result, the cardinality of the set is large and the optimization problem described in Equation \eqref{Def:optimal} is not computable through exhaustive search. However, in Section \ref{sec:exist} we show that
this optimization problem can be expressed in a computable form by restricting the search to a specific subset of stationary and memoryless strategies called \textit{threshold based strategies}. % which has cardinality of the continuum $|\mathbb{R}|$. 
 More precisely, we show that any optimal strategy is equivalent to a threshold based strategy where equivalence betweens strategies defined below

\begin{Definition}[\bf{Equivalence}]
\label{def:eq}
For the scheduling setup $(n,N_{max}, \underline{w}^n, \overline{w}^n, f_{R^m})$ two strategies $Q$ and $Q'$ are called equivalent ($Q\sim Q'$) if:
\begin{align*}
\lim_{t\to \infty} \frac{1}{t}\sum_{i=1}^t P\left(Q_i\big(R^{m\times i}\big)=Q'_i\big(R^{m\times i}\big)\right)=1.
\end{align*}
\end{Definition}

\begin{Definition}[\bf{Stationary and Memoryless}]
A strategy $Q=(Q_t)_{t\in \mathbb{N}}$ is called memoryless if $Q_t(R^{m\times t}), t\in \mathbb{N}$ is only a function of the performance vector $(R_{1,t}, R_{2,t}, \cdots, R_{m,t})$ corresponding to time $t$. For the memoryless strategy Q, we write $Q_t(R^m)$ instead of $Q_t(R^{m\times t})$ when there is no ambiguity. 
A memoryless strategy is called stationary if $Q_t(R^m)=Q_{t'}(R^m)$ for any $t,t'\in \mathbb{N}$. 
\end{Definition}

\begin{Lemma} \label{lem:convergence}
For a memoryless and stationary strategy $Q$, the following limits exist:
\begin{align}
A_{i}^Q &= \lim_{t \rightarrow \infty} A_{i,t}^Q,\qquad 
U^{Q}=\lim_{t \rightarrow \infty} U^Q_t.
\label{eq:temp-share-conv}
\end{align}
\end{Lemma}
The proof is provided in the Appendix.

\begin{Definition}[\bf{TBS}]
For the scheduling setup $(n,N_{max}, \underline{w}^n, \overline{w}^n, f_{R^m})$ a threshold based strategy (TBS) is characterized by the vector $\lambda^n\in \mathbb{R}^n$. The strategy $Q_{TBS}(\lambda^n)=(Q_{TBS,t})_{t\in \mathbb{N}}$ is defined as:
\begin{align}
Q_{TBS,t}\big(R^{m\times t}\big)=\argmax_{\mathcal{V}_j\in\mathsf{V}} ~S\big(\mathcal{V}_j,R_{t,j}\big), ~t\in \mathbb{N},
\label{Eq:thresh_str}
\end{align}
where $S\big(\mathcal{V}_j,R_{t,j}\big)=R_{t,j}+\sum_{i=1}^n \lambda_i \mathbbm{1}_{\{u_i\in\mathcal{V}_{j}\}}$ is the `scheduling measure' corresponding to the virtual user $\mathcal{V}_j$. 
 The resulting temporal shares are represented as $w_i=A_i^{Q_{TBS}}, i\in [n]$. The utility of the TBS is written as $U_{w^n}(\lambda^n)$. 
The space of all threshold based strategies is denoted by $\mathcal{Q}_{TBS}$.
\label{def:U_TBS}
\end{Definition}
%\textcolor{blue}{Farhad: One may wonder what happens if multiple virtual user take the highest scheduling measure. Shall we mention that the probability of tie is zero due to jointly continuity of performance values? Therefore we can randomly choose among the virtual users with highest scheduling measure. We can also mention that discrete and mixed distributions of performance value are studied in Section \ref{sec:discrete}. I THINK WE SHOULD MENTION THIS LATER IN THE DISCRETE SECTION}
% \begin{Definition}
% Consider the threshold based strategy $Q_{TBS}(\lambda^n)$.

% \end{Definition}

\begin{Remark}
The output of $Q_{TBS,t}, t\in \mathbb{N}$ in \eqref{Eq:thresh_str} depends only on the threshold vector $\lambda^n$ and the realization of $R^m$ at time t. As a result, the strategy is stationary and memoryless.  
\end{Remark}

\begin{Example}
Consider the NOMA system described in Example \ref{Ex:rep}. A TBS with threshold vector $\lambda^2=(\lambda_1,\lambda_2)$ has the following scheduling measures:
\begin{align*}
&S\big(\mathcal{V}_1,R_{t,1}\big)=R_{t,1}+\lambda_1,\\ &S\big(\mathcal{V}_2,R_{t,2}\big)=R_{t,2}+\lambda_2, \\ &S\big(\mathcal{V}_{1,2},R_{t,1,2}\big)=R_{t,1,2}+\lambda_1+\lambda_2.
\end{align*}
The virtual user with the highest scheduling measure is chosen at each time-slot. Note that the probability of a tie among the scheduling measures is 0 since the performance vector $R^m$ is assumed to be jointly continuous. 
% It is straightforward to verify that the temporal share of user $u_i$ is a strictly increasing function of $\lambda_i$. 
% We should continue example 1 by giving the table for $R_1+\lambda_1, R_2+\lambda_2, R_{1,2}+\lambda_1+\lambda_2, \cdots$ and explaining how the TBS works. 
\end{Example}

\section{Existence of Optimal Threshold Based Strategies}
\label{sec:exist}
In this section, we show that for any scheduling problem with temporal fairness constraints, optimal utility can be achieved using a threshold based scheduling strategy. Therefore, in considering the optimization problem described in Equation \eqref{Def:optimal}, the set of strategies $\mathcal{Q}$ can be restricted to the set of threshold based strategies. Furthermore, we show that any scheduling strategy which achieves optimal utility is equivalent to a threshold based strategy, where equivalence between strategies is defined in Definition \ref{def:eq}.

%\textcolor{blue}{Lemma:  we should add a lemma in the next section showing the fact that for any stationary and memoryless strategy, it can be shown that temporal shares and network utility would converge according to LLN}

% In this section, we show that for any scheduling problem with temporal fairness constraints, optimal utility can be achieved using a threshold based scheduling strategy. Therefor, in considering the optimization problem described by Equation \eqref{Def:optimal}, the set of strategies $\mathcal{Q}$ can be restricted to the set of threshold based strategies. Furthermore, we show that any scheduling strategy which achieves optimal utility is equivalent to a threshold based strategy, where equivalence between strategies is defined below.
% \begin{Definition}
% For the scheduling problem $(n,m, \underline{w}^n, $ $  \overline{w}^n, \mathcal{U}, \mathsf{V}, R^m)$ two strategies $Q$ and $Q'$ are called equivalent ($Q\sim Q'$) if:
% \begin{align*}
% \lim_{t\to \infty} \frac{1}{t}\sum_{i=1}^t P(Q_t(R^{m\times t})=Q'_t(R^{m\times t}))=1.
% \end{align*}
% \end{Definition}

\subsection{Optimal Temporally Fair NOMA Scheduling}
\label{subsec:exist}
 
%\begin{Theorem} 
%For the NOMA scheduling setup $(n,m, w^n, w^n, \mathcal{U}, \mathsf{V}, f_{R^m})$, assume that $\mathcal{Q}\neq \phi$. Then, there exists an optimal threshold based strategy $Q_{TBS}$. Furthermore, for any optimal strategy $Q$, there exists a threshold based strategy $\widehat{Q}$ such that $Q\sim \widehat{Q}$. 
%\label{th:eq:normal}
%\end{Theorem}
%\textcolor{blue}{combine Theorem 1 and 2. Add a remark and explain the importance of equivalency part of the theorem}
 %Theorem \ref{th:eq:normal} can be extended to fairness under inequality constraints as follows. 
\begin{Theorem} 
For the NOMA scheduling setup $(n,N_{max}, \underline{w}^n, \overline{w}^n, f_{R^m})$, assume that $\mathcal{Q}\neq \phi$. Then, there exists an optimal threshold based strategy $Q_{TBS}$. Furthermore, for any optimal strategy $Q$, there exists a threshold based strategy $Q'$ such that $Q\sim Q'$. 
\label{th:neq:normal}
\end{Theorem}
 The condition $\mathcal{Q}\neq \phi$ in Theorem \ref{th:neq:normal} is called the feasibility condition and is investigated in Section \ref{sec:feas}.  The proof of the theorem follows form the following steps:

 \begin{itemize}
 \item {Under equality temporal demand constraints where $\underline{w}^n=\overline{w}^n=w^n$, existence of optimal TBSs follows from a generalization of the intermediate value theorem called the Poincar\'{e}-Miranda Theorem \cite{kulpa1997poincare}.}
 \item {The uniqueness of an optimal strategy up to equivalence follows from a variant of the dual (Lagrangian multiplier) optimization method. }
 \item{Under inequality temporal demand constraints, the proof of existence follows by discretizing the feasible space and solving the optimization for each point on the discretized space under equality constraints.} 
 \end{itemize}
 The complete proof of Theorem \ref{th:neq:normal} is provided in the Appendix. The following corollaries follow from the proof.

\begin{Corollary}
Consider the NOMA scheduling setup under equality temporal demand constraints $(n,N_{max}, {w}^n, {w}^n, f_{R^m})$, where $\mathcal{Q}\neq \phi$. 
There exists a unique TBS $Q_{TBS}$ satisfying the temporal demand constraints. This unique TBS is the optimal scheduling strategy for this setup.
\label{cor:equal}
\end{Corollary}

The following corollary states that the search for the optimal strategy in equation \eqref{Def:optimal} may be restricted to the set of TBSs. 
\begin{Corollary}
\label{cor:two_step}
For the scheduling setup $(n,N_{max}, \underline{w}^n, \overline{w}^n, f_{R^m})$. The optimal achievable utility is given by:
\begin{align}
U^*_{\underline{w}^n, \overline{w}^n}\triangleq 
\max_{\lambda^n: \underline{w}_i\leq A_i^{Q_{TBS}}\leq \overline{w}_i} U_{w^n}(\lambda^n),
\label{eq:opti}
\end{align}
where $U_{w^n}(\lambda^n)$ is defined in Definition \ref{def:U_TBS}.
\end{Corollary}

\begin{Remark}
Note that from Theorem \ref{th:neq:normal}, $U^*_{w^n,w^n}=U_{w^n}(\lambda^n)$ since the threshold based strategy achieves optimal utility among all strategies with temporal shares equal to $w^n$.
\end{Remark}

In Section \ref{sec:sim}, we use the Corollary \ref{cor:CS} to provide a low complexity algorithm for constructing optimal TBSs.
\begin{Corollary}
For the NOMA scheduling setup $(n,N_{max}, \underline{w}^n, \overline{w}^n, f_{R^m})$. Assume that there exist positive thresholds $\lambda_1,\lambda_2,\cdots,\lambda_n$ satisfying the complimentary slackness conditions:
\begin{align*}
&\lambda_i\left(A^{Q_{TBS}}_i-\underline{w}_i\right)=0, \forall i\in [n],\\
&\underline{w}_i\leq A^{Q_{TBS}}_i\leq \overline{w}_i, \forall i\in [n],
\end{align*}
where $Q_{TBS}$ is the TBS corresponding to the threshold vector $\lambda^n$. Then, $Q_{TBS}$ is an optimal scheduling strategy. 
\label{cor:CS}
\end{Corollary}

\begin{Remark}
Note that the complementary slackness conditions in Corollary \ref{cor:CS} are only written in terms of the lower temporal demands. Similar sufficient conditions can be derived in terms of the upper temporal share demands.  
\end{Remark}
\begin{Remark}
If $Q \sim Q'$, then the two scheduling strategies activate the same subsets of users in almost all time-slots. As a result, the two strategies have the same performance under any long-term fairness and utility criteria. Let $\mathcal{Q}^*$ be the set of all optimality achieving strategies under temporal demand constraints. A consequence of Theorem \ref{th:neq:normal} is that all of the strategies in $\mathcal{Q}^*$ have the same performance with each other under any additional utility or fairness criteria. 
\end{Remark}

\subsection{Feasible Temporal Share Region}
\label{sec:feas}

Section~\ref{subsec:exist} affirms the existence of a TBS that achieves the optimal average system utility given that the temporal demands are feasible. However, some values of $(\underline{w}^n, \overline{w}^n)$, are not achievable by any scheduling strategy. In other words, the set $\mathcal{Q}$ is empty for certain pairs of constraint vectors $(\underline{w}^n, \overline{w}^n)$. In this section, %Theorem~\ref{Lem:feas}, 
 we provide a variable elimination method which allows us to characterize the feasible region for a given scheduling setup as a function of its temporal demand vectors.
%which are functions of a set of auxiliary variables representing the temporal share of each virtual user. We further characterize the feasible region as a system of linear inequalities which are functions of the temporal demand constraints by eliminating the auxiliary variables 
%in Theorem~\ref{Lem:feas}. While the feasibility of a given temporal share vector may be verified using the system of linear inequalities in Theorem~\ref{Lem:feas} through linear programming, e.g., with a dummy objective function, the system of inequalities after the variable elimination step may be directly verified without further complexity.% In the sequel, the explicit system of inequalities that are solely based on temporal demand constraints will be used to provide a visualization of the feasibility region in Example~\ref{ex:feas}. Such explicit characterization facilitates finding the optimal solution of the generalized problem under inequality constraints in Section \ref{sec:robbins-monro}. 

% In this section, we characterize the region of \textit{feasible} $(\underline{w}^n, \overline{w}^n)$ pairs for which $\mathcal{Q}\neq \phi$.
\begin{Definition}
A scheduling setup $(n,N_{max}, \underline{w}^n, \overline{w}^n, f_{R^m})$ is called feasible if $\mathcal{Q}\neq \phi$. The set of temporal shares $w^n$ for which the setup is feasible under equality constraints is called the feasible region and is denoted by $\mathcal{W}$. 
\end{Definition}
The following theorem characterizes the set of all feasible scheduling setups.
\begin{Theorem}
A scheduling setup with equality temporal constraints $(n,N_{max}, {w}^n, {w}^n, f_{R^m})$ is feasible iff there exist a set of positive values $\{a_{j}: j\in [m]\}$ satisfying the following bounds: 
\begin{align}
&  \sum_{j\in [m]} a_j= 1,\quad
\sum_{j: u_i\in \mathcal{V}_j} a_{j}= w_i,  \forall i\in [n],\label{eq:1}
\quad  0\leq a_{j}, \forall j\in [m].
\end{align}
Furthermore, the scheduling setup with inequality temporal constraints $(n,N_{max}, \underline{w}^n, \overline{w}^n, f_{R^m})$ is feasible iff there exist a set of positive numbers $\{w_{i}: i\in [n]\}$. Such that i) $(n,N_{max}, {w}^n,$ $ {w}^n, f_{R^m})$ is feasible, and ii) the following bounds are satisfied:
\begin{align*}
\forall i\in [n]: \underline{w}_i\leq w_i\leq \overline{w}_i.
\end{align*}
\label{Lem:feas}
\end{Theorem}
\textit{Proof Sketch:} In order to prove that $\mathcal{Q}\neq \phi$ under equality constraints, we use a WRR scheduler in which the weight assigned to $\mathcal{V}_j$ is equal to $a_j$. Under inequality temporal constraints, in order to have $\mathcal{Q}\neq \phi$, it suffices that there is a vector of temporal shares in the feasible region which satisfies the inequality constraints. 
\qedsymbol

The  feasibility conditions in Theorem \ref{Lem:feas} are written in terms of auxiliary variables $a_j, j\in [m]$ which can be interpreted as the temporal shares of the virtual users. However, it is often desirable to write the conditions in terms of the temporal share vector $w^n$. The conditions can be re-written in the desired form using the Fourier-Motzkin elimination (FME) algorithm. The standard FME algorithm has worst case computational complexity of order $O(m^{2^{m-n-1}})$ \cite{FME2}.  In \cite{FME_Farhad}, a method for variable elimination was proposed which has a significantly lower computation complexity.
The method leads to the following algorithm for determining the feasible region:
\\\textbf{Step 1.} Eliminate $a_{i}, i\leq n+1$ using the equality constraints \eqref{eq:1}:
\begin{align*}
&a_i= w_i-\sum_{j:u_i\in \mathcal{V}_j}a_j, i\in [n],\quad a_{n+1}= 1-\sum_{j\in [m],j\neq n+1} a_j.
\end{align*}
That is, replace $a_i, i\leq n+1$ in all inequality constraints \eqref{eq:1} by the right hand sides in the above equations.
\\\textbf{Step 2.} Define $c_{l,j}, l\in [m], 1\leq j \leq n $ as the coefficient of $w_j$ in the $l$th inequality. Also, define $c_{l,j}, l\in [m], n+2\leq j \leq m $ as the coefficient of $a_j$ in the $l$th inequality.
Construct the dual system of equations:
\begin{align*}
\sum_{l\in [m]}c_{l,j}x_l=0, \quad  x_l\in \mathbb{N}\cup\{0\}, ~ j\in \{n+2,n+3,\cdots,m\}.
\end{align*}  
\\\textbf{Step 3.} Use the  Normaliz algorithm \cite{Normaliz} to find the Hilbert basis for the solution space of the dual system. Let $\mathcal{B}= \{b_1^m.b_2^m,\cdots,b_k^m\}$ be the Hilbert basis, where $k$ is the number of Hilbert basis elements. 
\\\textbf{Step 4.} Let $b_i^m=(b_{i,1}, b_{i,2},\cdots,b_{i,m}), i\in [k]$. The following system of inequalities gives the feasible region:
\begin{align}
&0\leq \sum_{l\in [n+1]}\sum_{j\in [n]} b_{i,l} c_{j,l} w_j, ~ i\in [k].
\label{eq:FME}
\end{align}

The Elimination process is explained in the following example.
\begin{Example}
\label{ex:feas}
\begin{figure}
\centering \includegraphics[width=0.9\columnwidth, draft=false]{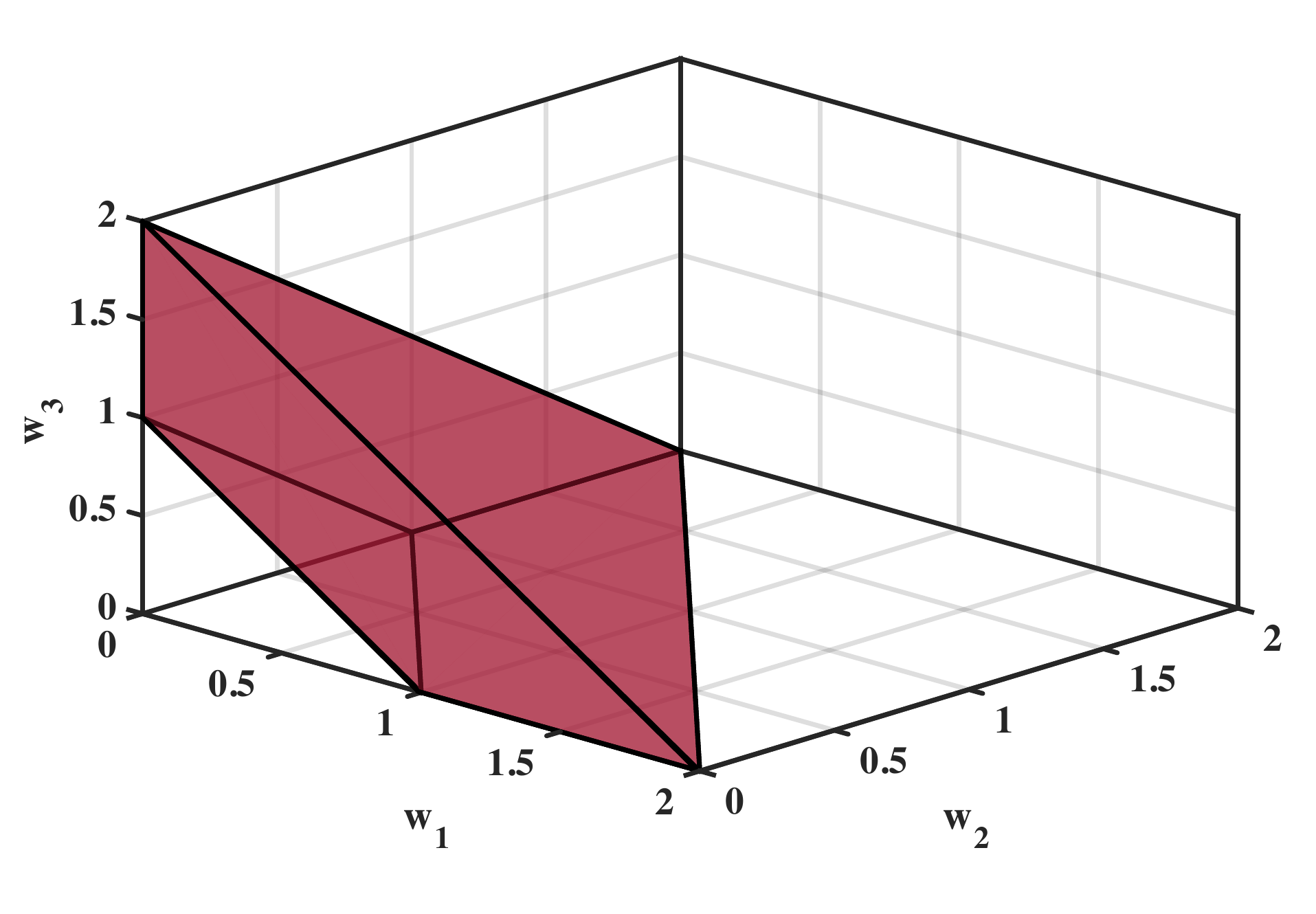}
\caption{The colored region shows the set of feasible weight vectors for the three user NOMA problem with $N_{max}=2$.}
\label{Fig:feas}
\end{figure}
Consider a three user downlink  NOMA scenario with $N_{max}=2$. The scheduling setup is feasible if there exist values $(a_1,a_2,a_3,a_{1,2},a_{1,3},a_{2,3})$ such that:
 \begin{align*}
 &0\leq a_1,~0\leq a_2,~0\leq a_3,~ 0\leq a_{1,2},~0\leq a_{1,3},~0\leq a_{2,3}\\ 
 &a_1+a_2+a_3+a_{1,2}+a_{1,3}+a_{2,3}=1,\\
 &a_1+a_{1,2}+a_{1,3}=w_1, \quad a_2+a_{1,2}+a_{2,3}=w_2, \\
 & a_3+a_{1,3}+a_{2,3}=w_3.
 \end{align*}
Using the equality constraints, one could write $a_1,a_2,a_3$ and $a_{1,2}$ as functions of $w_1,w_2,w_3, a_{1,3}$ and $a_{2,3}$. We get:
\begin{align*}
&0\leq 1+a_{1,3}-w_1-w_3,  \quad 0\leq 1+a_{2,3}-w_2-w_3, \\
&a_{1,3}+a_{2,3}\leq \min(w_3,w_1+w_2+w_3-1), \quad 0\leq a_{1,3}, \quad 0\leq a_{2,3} .
\end{align*}
There are a total of five inequalities, i.e. $k=5$. The dual system is given by:
\begin{align*}
&x_1-x_3+x_4=0,\qquad x_2-x_3+x_5=0.
\end{align*}
The Hilbert basis is $\mathcal{B}= \big\{(1,1,1,0,0), $ $ (0,0,1,1,1), $ $ (0,1,1,1,0), (1,0,1,0,1)\big\}$. From Equation \eqref{eq:FME}, the feasible region is as follows:
\begin{align*}
0\leq w_i\leq 1, ~i\in [3], \quad 1\leq w_1+w_2+w_3\leq 2.
\end{align*}
The region is shown in Figure \ref{Fig:feas}.

% Consider the NOMA scheduling problem in Example \ref{Ex:rep}. The scheduling problem is feasible if there exist values $(a_1,a_2,a_{12})$ such that:
% \begin{align*}
% &0\leq a_1,\quad 0\leq a_2,\quad 0\leq a_{12},\\
% &a_1+a_2+a_{12}=1,\\
% &a_1+a_{12}=w_1, a_2+a_{12}=w_2.
% \end{align*}
% Using the FME algorithm the conditions can be written as follows:
% \begin{align*}
% &0\leq w_1, \quad 0\leq w_2,\quad 1\leq w_1+w_2\leq 2.
% \end{align*}
% The feasible region is shown in Figure \ref{Fig:feas}.

\end{Example}

\begin{Remark}
Figure \ref{Fig:feas} can be interpreted as follows: since no more than two users can be activated at each time-slot, the total temporal share of all users cannot exceed two. The problem of determining the feasible region is more complicated for large NOMA systems where specific subsets of users cannot be activated simultaneously. In such instances the elimination method proposed above can be a valuable tool in determining the feasible region. 
\end{Remark}

% Lemma \ref{Lem:feas} provides conditions for feasibility of scheduling problems under equality constraints. The result can be extended as follows to problems with inequality constraints.

% \begin{Lemma}

% \label{Lem:feas2}
% \end{Lemma}
%\subsection{Complementary Slackness}
\section{Construction of Optimal Scheduling Strategies}
\label{sec:robbins-monro}
In the previous section, it was shown that for any feasible scheduling problem, optimal utility is achieved using TBSs. %Theorem \ref{th:neq:normal} provides existence results for optimal TBSs. However, the question of how to construct optimal scheduling strategies has remained unanswered. In this section,  .
In this section, we address the construction of optimal scheduling strategies and provide an iterative method which builds upon the Robbins-Monro algorithm \cite{robbins1951stochastic} to find optimal thresholds for TBSs. As mentioned in Remark \ref{Rem:perf}, the scheduler does not have access to the statistics of the performance vector. The algorithm proposed in this section uses the empirical observations of the realizations of the performance vector to find the optimal thresholds iteratively.

%Details are provided for constructing temporal fair strategies under equality constraints (i.e. when $\underline{w}^n=\overline{w}^n=w^n$). A heuristic algorithm for scheduling problems with inequality constraints is provided in the Section \ref{sec:sim}. 

\subsection{The Robbins-Monro Algorithm}
The Robbins-Monro algorithm \cite{robbins1951stochastic} is a method for finding the roots of the univariate function $f: \mathbb{R} \rightarrow \mathbb{R}$ based on a limited number of noisy samples of $f(x), x\in \mathbb{R}$. More precisely, assume that we take $l$ noisy samples $g_t=f(x_t)+\epsilon_t$  of the function $f(\cdot)$ at $x_t, t\in [l]$, , where the random variable $\epsilon_t$ is the sampling noise at time $t$ and $l$ is a fixed natural number. 
%Let $g_t(x_t)=f(x_t)+\epsilon_k, t\in [l], x_t\in \mathbb{R}$ be the noisy sample
The objective is to approximate the solution of $f(x)=0$ by choosing $x_t, t\in [l]$ suitably such that the approximation converges to the root as $l\to \infty$. 
%Robbins and Monro provided sufficient conditions for convergence of the sequence $x_{t+1}=x_t-s_tg_t$ converges to the solution of $f(x)=0$
%provided that the following conditions hold:
% \begin{enumerate}
% \item \textbf{Solvability:} The function $f(\cdot)$ has a root $x^*$.
% \item \textbf{Local Monotonicity:} $(x-x^*)f(x)>0, \forall x\not=x^*$.
% \item \textbf{Zero mean and i.i.d. noise:} $\epsilon_t, t\in \mathbb{N}$ is an i.i.d. sequence and $\mathbb{E}(\epsilon_t)=0$.
% \item \textbf{Step-size constraints:} $s_t>0$, $\lim_{t\rightarrow \infty} s_t=0$, $\sum_{t=1}^\infty s_t=\infty$, $\sum_{t=1}^\infty s^2_t<\infty$.
% \end{enumerate}
% The sequence $\{s_t\}_{t\in \mathbb{N}}$, is called the sequence of step-sizes. 
The algorithm was extended to find roots of multi-variate functions by Ruppert \cite{ruppert1985newton}.  Let $f: \mathbb{R}^n \rightarrow \mathbb{R}^n$ be a mapping of the $n$-dimensional Euclidean space onto itself. Let $g_t=f(x^n_t)+\epsilon^n_t$ be the noisy sample of $f(x_t^n)$ at $x^n_t, t\in [l]$. It can be shown that sequence $x^n_{t+1}=x^n_t-s_tg^n_t$ converges to the solution of the system $f(x^n)=0$ if the following conditions hold:
\begin{enumerate}
\item \textbf{Solvability:} The function $f(\cdot)$ has a root $x^{* n}$.
\item \textbf{Local Monotonicity:} $(x^n-x^{* n})^Tf(x^n)>0$, .
\item \textbf{Zero-mean and i.i.d. noise:} $\epsilon^n_t, t\in \mathbb{N}$ is an i.i.d. sequence and $\mathbb{E}(\epsilon^n_t)=0$.
\item \textbf{Step-size constraints:} $s_t>0$, $\lim_{t\rightarrow \infty} s_t=0$, $\sum_{t=1}^\infty s_t=\infty$, $\sum_{t=1}^\infty s^2_t<\infty$.
\end{enumerate}
\subsection{Finding Thresholds under Equality Constraints} \label{sec:RM-equality} \label{subsec:RM-equality}
In Corollary \ref{cor:equal}, we showed that any threshold based strategy satisfying the equality temporal  constraints is optimal. As a result, the objective of finding the optimal scheduling strategy is reduced to finding the thresholds which lead to a TBS satisfying the temporal demand constraints. 
More precisely, we are interested in finding the threshold vector $\lambda^n$ such that 
\begin{align}
A_i^{Q_{TBS}(\lambda^n)}=w_i, i\in[n].
\label{eq:non_lin}
\end{align}

Hence, finding the optimal TBS is equivalent to solving the non-linear system of equations in \eqref{eq:non_lin}. We show that the empirical observation of the realizations of $R^m$ at the BS is sufficient to find the optimal thresholds using the multi-variate version of the Robbins-Monro stochastic approximation method  \cite{ruppert1985newton}.

In the scheduling problem, consider $f_i(\lambda^n)=A_i^Q(\lambda^n)-w_i, i\in[n]$. We are interested in finding the root of the function $f(\cdot)$ provided that the root exists (i.e. $\mathcal{Q}\neq \phi$). We will show that conditions 1-4 provided above are satisfied for a suitable step-size sequence. 
Note that $A_i^Q(\lambda^n)$ depends on the statistics of the performance vector $R^m$ and is not explicitly available in practice. Assume that at time $t$, the scheduler uses the threshold vector $\lambda_t^n$. Then, it observes $g_{t,i}\triangleq \mathbbm{1}{\{u_i\in Q_t(\lambda^n_t)\}}-w_i$ which is a noisy sample of  $f(\lambda_t^n)= A_i^Q(\lambda_t^n)-w_i$. The sampling noise is $\epsilon_{t,i}=g_{t,i}(\lambda^n_t)-f_{i}(\lambda^n_t)=\mathbbm{1}{\{u_i\in Q_t(\lambda^n_t)\}}-A_i^Q(\lambda^n_t)$. The sequence of sampling noise vectors $\epsilon^n_t, t\in \mathbb{N}$ is an i.i.d sequence since $Q_t(\lambda^n_t)$  depends only on the realization of $R^m$ at time $t$. Furthermore, it is straightforward to show that $\mathbb{E}(\epsilon_{t,i})=0$. Therefore,
conditions $(1)$ and $(3)$ are satisfied. 
The following theorem shows that conditions $(2)$ and $(4)$ hold. 
\begin{Theorem}
The convergence conditions $(1)$-$(4)$ in the multi-variate Robbins-Monro algorithm are satisfied for step-size the sequence $s_t=\frac{1}{t}, t\in \mathbb{N}$. 
\label{th:converge}
\end{Theorem}

The proof is provided in the Appendix.

%One needs to only check condition (2). The condition follows by the fact that $A_{i}^Q$ is increasing in $\lambda_i$. The proof is omitted due to space limitations.  (WILL BE ADDED)

\subsection{Finding Thresholds under General Temporal Constraints}
In this section, we provide an algorithm for finding the optimal thresholds for TBSs under general temporal demand constraints. The optimization algorithm uses a combination of the gradient projection method \cite{bertsekas1999nonlinear} and the Robbins-Monro algorithm described in the previous section. In order to use the gradient projection method, the scheduler needs to know $\nabla U^*_{{w}^n,{w}^n}$. As mentioned in Remark \ref{Rem:perf}, the scheduler does not know the statistics of the performance vector. As a result, the scheduler must estimate  $\nabla U^*_{{w}^n,{w}^n}$ using a gradient estimation method based on the empirical observations of the performance vector. However, estimating the gradient has high computational complexity. We propose a heuristic algorithm which does not require estimating  the gradient. This heuristic algorithm is used in Section \ref{sec:sim} for simulations. 
%is broken into two steps. First, the algorithm finds the temporal weights corresponding to the optimal scheduler using the . Next, it finds the optimal thresholds for the given temporal weights using the .  

We build upon the observation in Corollary \ref{cor:two_step} which shows that the optimal utility $U^*_{\underline{w}^n, \overline{w}^n}$ can be found by maximizing $U_{w^n}(\lambda^n)$ over all feasible TBSs. 
In the next lemma, we show that the function $U^*_{\underline{w}^n, \overline{w}^n}$  is jointly concave in $(\underline{w}^n, \overline{w}^n)$. Consequently, the optimization in \eqref{eq:opti} can be performed using standard gradient projection methods \cite{bertsekas1999nonlinear}. Note that the gradient projection method requires prior knowledge of the feasible set. The feasible set is characterized using the method in Section \ref{sec:feas}.

\begin{Lemma}
The optimal achievable utility $U^*_{\underline{w}^n, \overline{w}^n}$ is jointly concave as a function of $(\underline{w}^n, \overline{w}^n)$. 
\label{lem:concave}
\end{Lemma}
The proof is provided in the Appendix. 
Algorithm \ref{alg:two-stage} describes a method for finding the optimal thresholds under inequality temporal constraints. The algorithm performs an iterative two-step optimization. At each iteration, in the first step, given a fixed temporal demand vector $w^n$, the Robbins-Monro algorithm is used to find the thresholds  under equality temporal constraints. Next, the gradient projection method is used to update the weight vector $w^n$ based on $\nabla U_{w^n,w^n}$. The algorithm converges to the optimal utility due to the concavity of $U^*_{\underline{w}^n, \overline{w}^n}$. The iteration stops if $\Delta\geq\epsilon$, where $\Delta$ represents the variation in $w^n$ at each step and $\epsilon$ is the stopping parameter. 

\begin{algorithm}[h]
\caption{Two-stage Threshold Optimization in TBS}
\small
\begin{algorithmic}[1]
\STATE Obtain feasibility region $\mathcal{W}$ (Section \ref{sec:feas}).
\STATE Set $\epsilon>0$ and $\Delta=2\epsilon$. 
\STATE Choose initial demand vector $w^n=w_0^n$. 
    \WHILE{$\Delta\geq \epsilon$}
    	\STATE Find  $U^*_{{w}^n,{w}^n}$ and corresponding $\lambda^n$ (Robbins-Monro Algorithm in Section \ref{subsec:RM-equality}).
    	\STATE Update $w^n$ based on $\nabla U^*_{{w}^n,{w}^n}$ (gradient projection step).
        \STATE $\Delta=||w^n_{new}-w^n||$
 %       \STATE $w^n \leftarrow w^n_{new}$
    \ENDWHILE
  \end{algorithmic}
  \label{alg:two-stage}
\end{algorithm}

% \begin{Remark}
  
% %In Section \ref{sec:sim} we provide a low complexity heuristic algorithm based on the complimentary slackness conditions which does not require knowledge of $\nabla U^*_{{w}^n,{w}^n}$. 
% \label{rem:CS}
% \end{Remark}

\begin{Remark}
The choice of the initial demand vector $w_0^n$ does not affect convergence since $U^*_{{w}^n,{w}^n}$ is jointly concave as shown in Lemma \ref{lem:concave}.
\end{Remark}

Algorithm \ref{alg:two-stage} requires estimating the gradient of $U^*_{{w}^n,{w}^n}$ which entails high computational complexity. As an alternative, we propose Algorithm \ref{alg:robbins-monro} which is
a low complexity heuristic variation of Algorithm \ref{alg:two-stage}.
The algorithm is constructed using the complementary slackness conditions provided in Corollary \ref{cor:CS}. Algorithm \ref{alg:robbins-monro}  replaces gradient projection by a simple perturbation step. This algorithm is used in Section \ref{sec:sim} for simulations.

The algorithm starts with a vector of initial thresholds. At time-slot $t$, it chooses virtual user $\mathcal{V}_t$ to be activated based on the threshold vector $\lambda_t^n$.  It updates the temporal shares and thresholds based on the scheduling decision at the end of the time-slot (line 2-5). The update rule for the thresholds given in line 5 is a variation of the Robbins-Monro update described in Section \ref{sec:robbins-monro}. The parameter $s$ is the step size. Lines (6-13) verify that the temporal demand constraints and dual feasibility conditions are satisfied. The computational complexity of the algorithm at each time-slot is proportional to the number of virtual users and is $O(n^2)$.

%We set parameter $\delta$ and $\Delta$ to  $=0.001$ when using Algorithm \ref{alg:robbins-monro}.  \textcolor{blue}{mention that this algorithm can also resolve tie situations}

\begin{algorithm}[h]
\caption{Heuritic Threshold Optimization in TBS }
\textbf{Initialization}: $\lambda_{1,i}=0, i \in [n]$
\small
  \begin{algorithmic}[1]
    \FOR {$t \in \mathbb{N}$}
    	\STATE $\mathcal{V}_t=Q_t(\lambda^n_t)$
        \STATE $A^Q_{t+1,i}=A^Q_{t,i}+\frac{1}{t+1}\Big(\mathbbm{1}{\{i\in\mathcal{V}_t\}}-A^Q_{t,i}\Big)$
    	\STATE $\lambda_{min}=\min_{i\in [n]} \lambda_{t,i}$
        \STATE $\lambda_{t+1,i}=\lambda_{t,i}-s\Big(\lambda_{t,i}-\lambda_{min}\Big)\Big(\mathbbm{1}{\{u_i\in\mathcal{V}_t\}}-\underline{w}_i\Big), i\in [n]$
        \FOR{$i=1$ to $n$}
        	\IF{$\lambda_{t,i}=\lambda_{min}$ and $A^Q_{t+1,i}<\underline{w}_i$ }
            	\STATE $\lambda_{t+1,i}=\lambda_{t,i}+s\Big(\underline{w}_i-A^Q_{t+1,i}\Big)$
            \ENDIF
            \IF{$\lambda_{t,i}=\lambda_{min}$ and $\lambda_{min}<0$}
            	\STATE $\lambda_{t+1,i}=\lambda_{t,i}+s$
            \ENDIF
        \ENDFOR
    \ENDFOR
  \end{algorithmic}
  \label{alg:robbins-monro}
\end{algorithm}

\section{Discrete and Mixed Performance Variables} \label{sec:discrete}
So far, we have assumed that the performance vector $R^m$ is jointly continuous. The proofs provided for Theorems \ref{th:neq:normal} and \ref{th:converge} rely on the joint continuity assumption. However, in practice this is not usually the case. In most practical scenarios, the performance vector is a vector of discrete or mixed random variables. For instance,  in cellular systems, the performance vector is discrete due to the use of discrete modulation and coding schemes. 

%The randomness in the quality of wireless channels is due to several factors. The large-scale and small-scale fading effects result in a stochastic SINR for each user's channel. 
The performance value of a virtual user is a function of the SINR of its elements. For instance, in Example \ref{Ex:NOMA}, sum-rate was considered as the performance value which is a logarithmic function of the SINRs.
%The SINR is modeled as a continuous random variable. 
Since the logarithm function is continuous, the performance vector is a jointly continuous vector of random variables. 
However, in practical scenarios, the function which relates the SINR to the performance value is neither injective nor continuous. The function is determined by the choice of the modulation and coding schemes at each time-slot. % In such scenarios, the preformance value is a staircase function of the SINR which maps different intervals of SINR to different rates according to the corresponding modulation and coding rates. This leads to a discrete probability distribution on the performance values.
Moreover, in some applications, the performance value is approximated by a truncated Shannon rate function, i.e. $R=\min\{\log_2(1+\text{SINR}),\gamma_{max}\}$, where $\gamma_{max}$ is the maximum data rate supported by the system. In this case, the performance value has a mixed distribution function.
Figure \ref{fig:ecdf-performance} shows the empirical CDF of the performance value of virtual user $\mathcal{V}_{1,2}$ in the downlink of a two user NOMA system with SIC, where the performance value is taken as i) the Shannon sum-rate as discribed in Example \ref{Ex:NOMA}, ii) truncated Shannon sum-rate, and iii) sum-rate with LTE modulation and coding schemes. In the truncated Shannon sum-rate model we use $\gamma_{max}=4$ bps/Hz and in the LTE rate model we use the parameters provided in \cite[Table 7.2.3-1]{3gpp-rate}, where 15 combinations of modulation and coding schemes are used. 

\begin{figure}[t]
 \centering \includegraphics[width=0.75\linewidth, draft=false]{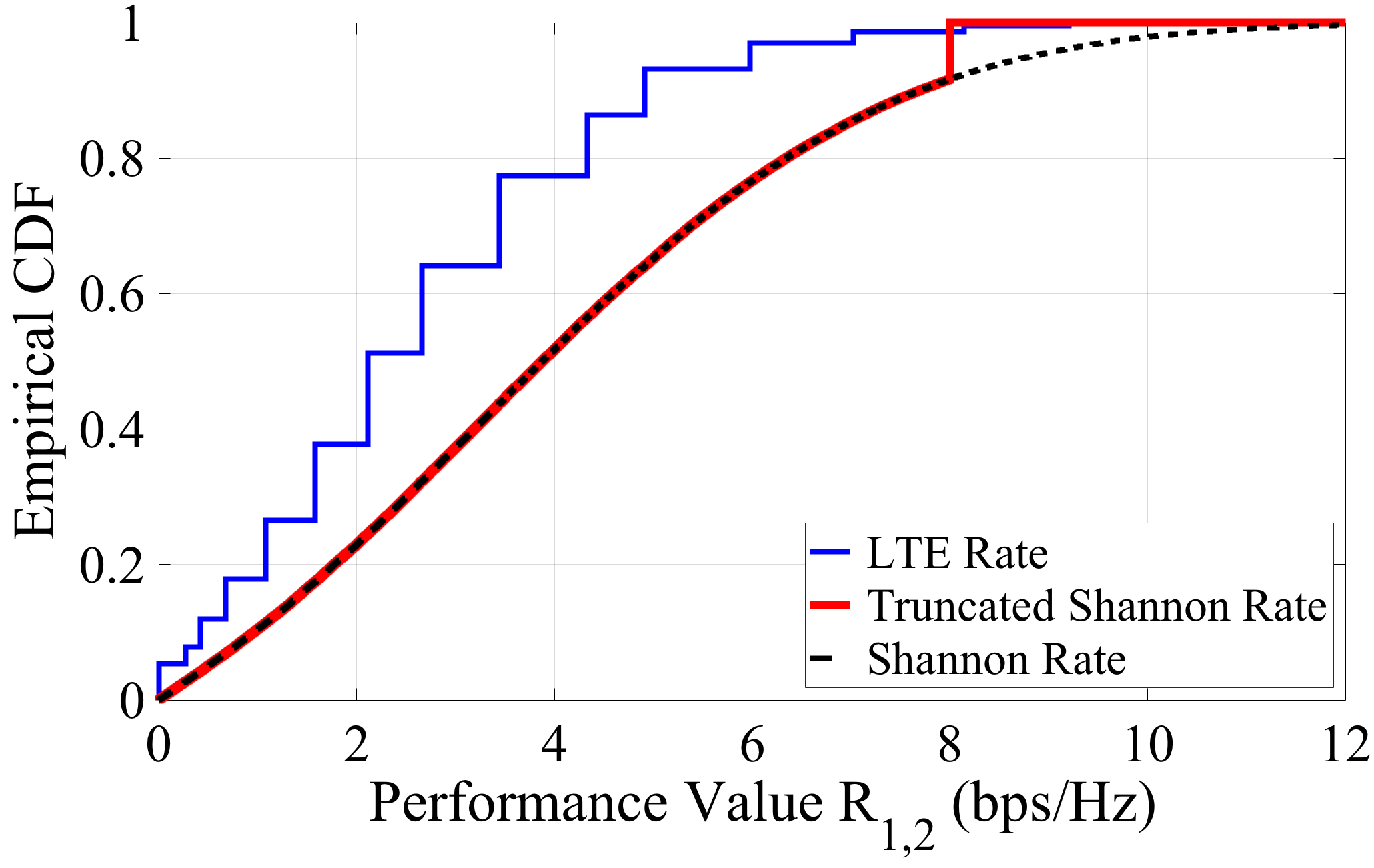}
 \caption{Empirical CDF of performance value of virtual user $\mathcal{V}_{1,2}$ in two user NOMA.}
 \label{fig:ecdf-performance}
\end{figure} 

The analysis provided in the previous sections cannot be applied  directly  to mixed and discrete performance vectors. The reason is that when the performance values are jointly continuous, the probability of having a tie among the scheduling measures in a TBS is zero. %Consequently, the temporal shares are not affected by the tie-breaking decision rule by the scheduler.
However, there may be more than one virtual user with the highest scheduling measure in the case of a mixed or discrete performance vector. In such scenarios, there is a need for a tie-breaking rule which affects the optimality of the scheduler. One widely used solution in OMA scheduling is to use a stochastic tie-breaker where in the event of a tie, one of the users is activated randomly based on a given probability distribution called the tie-breaking probability \cite{liu-jsac}. This requires a joint optimization of the thresholds and the tie-breaking rule.

%solving an additional optimization problem to find the best tie-breaking decision rule.
%The complexity of this optimization grows exponentially in  the number of virtual users $m$. 

We propose a new method to handle mixed and discrete random variables in which the performance vector is perturbed by a small additive uniformly distributed noise vector. It is straightforward to show that the perturbed vector is jointly continuous. In particular, fix $l\in \mathbb{N}$. Define $\widetilde{R}_{t,j}(l)=R_{t,j}+N_{t,j,l}$, $j\in[m]$ where $N_{t,j,l}\sim Unif[-\frac{1}{l}, \frac{1}{l}]$. The variables $N_{t,j,l}$ are jointly independent. A TBS is designed for this new performance vector as described in Section \ref{sec:robbins-monro}. This TBS is then used to schedule for the original discrete system.
%Since the perturbed performance vector is jointly continuous, the existence of an optimal TBS and convergence of the Robbins-Monroe algorithm follow from the previous sections. 
The following theorem shows that the average system utility for the designed TBS approaches the optimal utility of the original system as $l \to \infty$.

\begin{Definition}
A discrete scheduling setup is characterized by $(n,N_{max}, \underline{w}^n, \overline{w}^n, R^m)$, where $R^m$ may be a discrete or mixed vector of random variables. 
\end{Definition}
\begin{Theorem}
$(n,N_{max}, \underline{w}^n, \overline{w}^n, f_{R^m})$
Let $Q^*$ be the optimal scheduling strategy for the setup $\Omega_0=(n,N_{max}, \underline{w}^n, \overline{w}^n, R^m)$. Let ${Q}_{\frac{1}{l}}$ be the optimal scheduling strategy for the setup $\Omega_{\frac{1}{l}}=(n,N_{max}, \underline{w}^n, \overline{w}^n, R^m+N_{\frac{1}{l}}^m)$, where  $N_{\frac{1}{l}}^m$ is a vector of independent $Unif[-\frac{1}{l},\frac{1}{l}]$ variables. Define $U^*$ and $U_{\frac{1}{l}}$ as the average system utility from $Q^*$ and $Q_{\frac{1}{l}}$ when applied to system $\Omega_0$, respectively. Then,
\begin{align*}
\lim_{l \to \infty} (U^*-U_{\frac{1}{l}})=0,
\end{align*}
where convergence is in probability. In other words, the utility of $Q_{\frac{1}{l}}$ when applied to setup $\Omega_0$  converges to the optimal utility  as $l\to \infty$.
\label{th:discrete}
\end{Theorem}

The proof is provided in the Appendix.

\begin{Remark}
The performance of the perturbed scheduler converges to the optimal performance as $l\to \infty$. % As a result, it is beneficial to take $l$ as large as possible.
However, the precision supported by the scheduler's equipment sets an upper limit on $l$.
\end{Remark}

%\begin{Remark}
%An interpretation of the method proposed above is that adding a small Gaussian noise places a stochastic tie-breaking rule on the scheduling policy. The difference is that the scheduler optimizes both the TBS and the tie-breaking rule in one step by finding the optimal thresholds for the perturbed system rather than explicitly optimizing the tie-breaking rule separately. 
%\end{Remark}

\section{Numerical Results and Simulations} \label{sec:sim}
\begin{table}[t]
\centering
\caption{Simulation parameters}
\scalebox{1}{
\begin{tabular}{ll} \toprule
$\bf Parameter$ 				 & $\bf Value$             \\ \midrule
Scenario                         & NOMA downlink with SIC \\
Number of users                  & $2,3,4,5,6$                      \\
Cell radius        				 & $100$ m                \\
System Bandwidth      					 & $10$ MHz                \\
Number of time-slots             & $5 \times 10^6$            \\
Maximum spectral efficiency      & $6$ bps/Hz            \\
Noise spectral density     	     & $-174$ dBm/Hz           \\
Noise figure                     & $9$ dB \\ 
Shadowing standard deviation     & $8$ dB                  \\
Pathloss in dB     & $128.1+37.6\log_{10}(d~\text{in km})$  \\ 
Rate in bps/Hz         & $\min\left\{\log_2(1+\text{SINR}),6\right\}$\\
User mobility model & Static, 2D random walk\\
\bottomrule
\end{tabular}
}
\label{tab:sim-param}
\end{table}

In this section, we provide various numerical examples and simulations to evaluate the performance of the approaches proposed in Sections \ref{sec:robbins-monro} and \ref{sec:discrete}. We simulate the DL of a small-cell NOMA system consisting of a BS and a number of users distributed uniformly at random in a ring around the BS with inner and outer radii of 20 m and 100 m, respectively. Table \ref{tab:sim-param} lists the network parameters. We consider $N_{max}=2$, i.e. an individual user or a pair of users is scheduled at each time-slot. We assume that there are no upper temporal demand constraints.
The user SINRs are modeled as described in Example \ref{Ex:NOMA} and the network utility is assumed to be truncated Shannon sum-rate. At each time-slot prior to scheduling, a max-min power optimization is performed for each virtual user \cite{liu2016fairness}. For a given virtual user, we find the transmit power which maximizes the minimum individual user rates in that virtual user. This max-min optimization allows for a balanced rate allocation within the virtual user. 
It can be shown that the max-min optimization is quasi-concave. Consequently, quasi-concave programming methods such as bisection search can be used to find the optimal transmit powers \cite{liu2016fairness}. Maximum BS transmit power constraint is chosen such that the average SNR of $10$ dB is achievable when a single user is active on the boundary of the cell. We use Algorithm \ref{alg:robbins-monro} described in Section \ref{sec:robbins-monro} for simulations. The step size $s$ is taken to be $0.001$.

\subsection{Performance Evaluation}
We evaluate the performance of the NOMA scheduler in a scenario where the users are static. As a benchmark, we consider an OMA system where a single user is activated at each time-slot. To find a temporal fair scheduler in an OMA system, we consider the setup when $N_{max}=1$. We also consider Round Robin (RR) scheduling as another benchmark. Figure \ref{fig:ecdf} shows the empirical cumulative distribution function (CDF) of the network throughput in a network with $n=5$ users and $\underline{w}_i=0.2, \forall i\in [5]$ for various scheduling strategies. We observe that there are significant improvements in terms of network throughput when the TBS (labeled opportunistic NOMA) is used compared to OMA (labeled opportunistic OMA) as expected. Furthermore, we note that RR scheduling in NOMA leads to a significant performance loss. 
%This shows that opportunistic scheduling leads to considerable gains in practical scenarios. 
The RR strategy chooses the virtual user regardless of the performance in that time-slot. 
As a result, the strategy is particularly inefficient in NOMA systems. The reason is that SIC which is used in NOMA may have a poor performance for a given virtual user in some time-slots. 

%since all the virtual users are scheduled in a RR manner regardless of the performance value while some of them have low performance value due to intense interference.  

Table \ref{tab:noma-gain} lists the percentage of the average throughput gain when using opportunistic NOMA scheduler compared to an opportunistic OMA scheduler. For a given number of users $n\in \{2,3,\cdots,6\}$, we simulated 100 independent realizations of the network. It can be observed that increasing the number of users boosts the NOMA performance gain. This is due to the fact that the number of virtual users increases as $n$ becomes larger. As a result, the NOMA scheduler has more options in the choice of the active virtual user.
\begin{figure}[t]
 \centering \includegraphics[width=0.8\linewidth, draft=false]{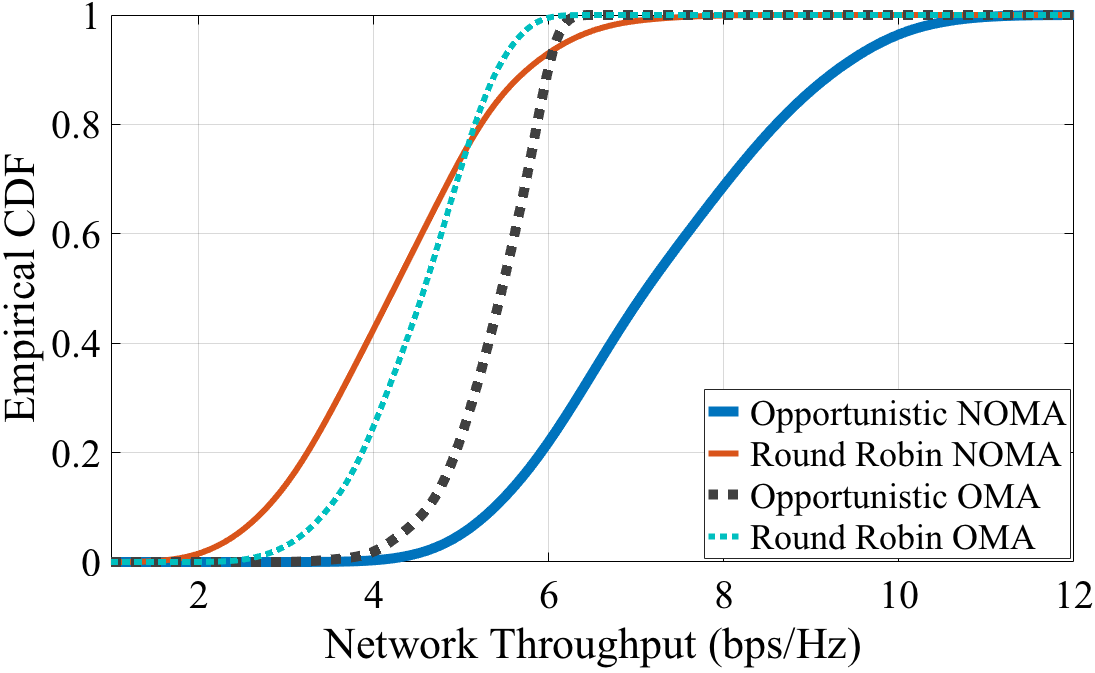}
  \caption{Empirical CDF of network throughput for NOMA and OMA systems using different scheduling schemes.}
%  \caption{Empirical CDF of network throughput for NOMA and OMA using the proposed scheduling algorithm.}
 \label{fig:ecdf}
\end{figure} 
\begin{table}[t]
\centering
\caption{Throughput gain of opportunistic NOMA scheduling over opportunisitc OMA scheduling }
\scalebox{1}{
\begin{tabular}{lccccc} \toprule
\textbf{Number of users}         & $2$  & $3$  & $4$  & $5$  & $6$    \\ \midrule
\textbf{Throughput gain} ($\%$)  & $7.33$  & $18.99$  & $27.08$  & $37.14$  & $45.47$  \\ 
\bottomrule
\end{tabular}
}
\label{tab:noma-gain}
\end{table}
\subsection{Convergence}
In this section, we investigate the evolution of the scheduling thresholds when running Algorithm \ref{alg:robbins-monro}. We consider a scenario with $n=5$ users and $\underline{w}^5=[0.1,0.1,0.4,0.3,0.1]$. We assume that the users' locations are fixed in the simulation time span. Figure \ref{fig:threshold}(a) shows the long-term user temporal shares $A^i_Q, t\in [5]$ and the user thresholds. It can be seen that the desired temporal demand constraints are satisfied. Moreover, the thresholds satisfy the optimality conditions discussed in Corollary \ref{cor:CS}. Figure \ref{fig:threshold}(b) shows the evolution of the thresholds in different iterations of Algorithm \ref{alg:robbins-monro}. %We observe that the thresholds approach their long-term values when using Algorithm \ref{alg:robbins-monro}. 

In Figure \ref{fig:temp-share}, we consider the previous scenario with mobile users. We model the mobility of the users by a two-dimensional random walk where each user takes one step per time-slot in a direction $\theta$ uniformly distributed in $[0,2\pi]$. Furthermore, we assume that the speed of the user at each step is randomly distributed between 1 m/s and 10 m/s. It is also assumed that the users do not exit the cell. In this scenario, the performance vector $R^m$ is not stationary and $A^Q_i(\lambda^n)$ is time-varying. Therefore, the optimal thresholds change over time. Figure  \ref{fig:temp-share} shows the long-term temporal share of the users and the evolution of the thresholds in time. It can be seen that the thresholds track the variation of the system and the desired temporal demand constraints are satisfied.
% \begin{figure}[t]
%  \centering \includegraphics[width=0.6\linewidth, draft=false]{Temporal_shares_v3.png}
%  \caption{Temporal share of the users compared to their minimum temporal weights.}
%  \label{fig:temp-share}
% \end{figure} 
% \begin{figure}[t]
%  \centering \includegraphics[width=0.6\linewidth, draft=false]{Thresholds_v3.png}
%  \caption{The evolution of scheduling thresholds in time when using Algorithm \ref{alg:robbins-monro}. The horizontal axis is the sampled time-slot index where the sampling parameter $H$ is set to $0.1$.}
%  \label{fig:threshold}
% \end{figure} 

%  \begin{figure}
%     \centering
%         \begin{subfigure}[h]{0.65\linewidth}
%             \includegraphics[width=\linewidth, draft=false]{Temporal_shares_v3.png}
%             \caption[]
%             {{\small }}    
%             \label{fig:temp-share}
%         \end{subfigure}

%         \begin{subfigure}[h]{0.65\linewidth} 
%         \includegraphics[width=\linewidth, draft=false]{Thresholds_v3.png}
%             \caption[]
%             {{\small}}    
%             \label{fig:threshold}
%         \end{subfigure}
%         \caption{(a) Temporal share of the users compared to their minimum temporal weights when the location of the users are fixed, (b) The evolution of scheduling thresholds in time when using Algorithm \ref{alg:robbins-monro}. The horizontal axis is the sampled time-slot index where the sampling parameter $H$ is set to $0.1$.}
% \end{figure}
 \begin{figure}[h]
    \centering
        \includegraphics[width=0.82\linewidth, draft=false]{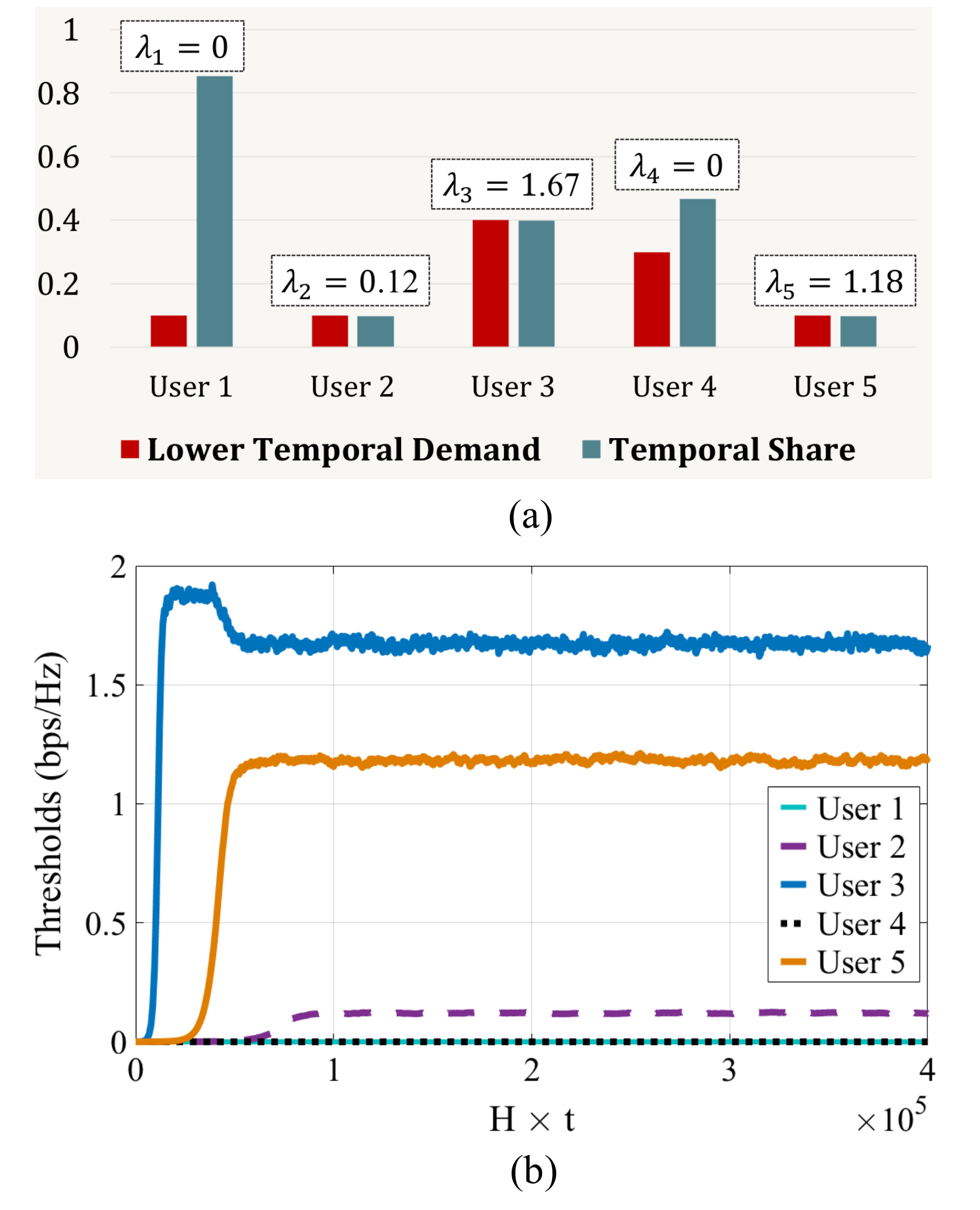} 
        \caption{(a)  Long-term temporal share and the lower temporal demands of the users, (b) The evolution of scheduling thresholds in Algorithm \ref{alg:robbins-monro}. The horizontal axis is the sampled time-slot index, where the sampling parameter $H$ is set to $0.1$.}
        \label{fig:threshold}
        \vspace{-4mm}
\end{figure}

% \begin{figure}
%     \centering
%         \begin{subfigure}[h]{0.65\linewidth}
%             \includegraphics[width=\linewidth, draft=false]{Temporal_shares_mobile_v3.png}
%             \caption[]
%             {{\small }}    
%             \label{fig:temp-share}
%         \end{subfigure}
%         \begin{subfigure}[h]{0.65\linewidth} 
%         \includegraphics[width=\linewidth, draft=false]{Thresholds_mobile.pdf}
%             \caption[]
%             {{\small}}    
%             \label{fig:threshold}
%         \end{subfigure}
%         \caption{(a) Temporal share of the users in a mobile scenario, (b) The evolution of scheduling thresholds in time. The sampling parameter $H$ is set to $0.1$.}
% \end{figure}
\begin{figure}[h]
    \centering
            \includegraphics[width=0.82\linewidth, draft=false]          {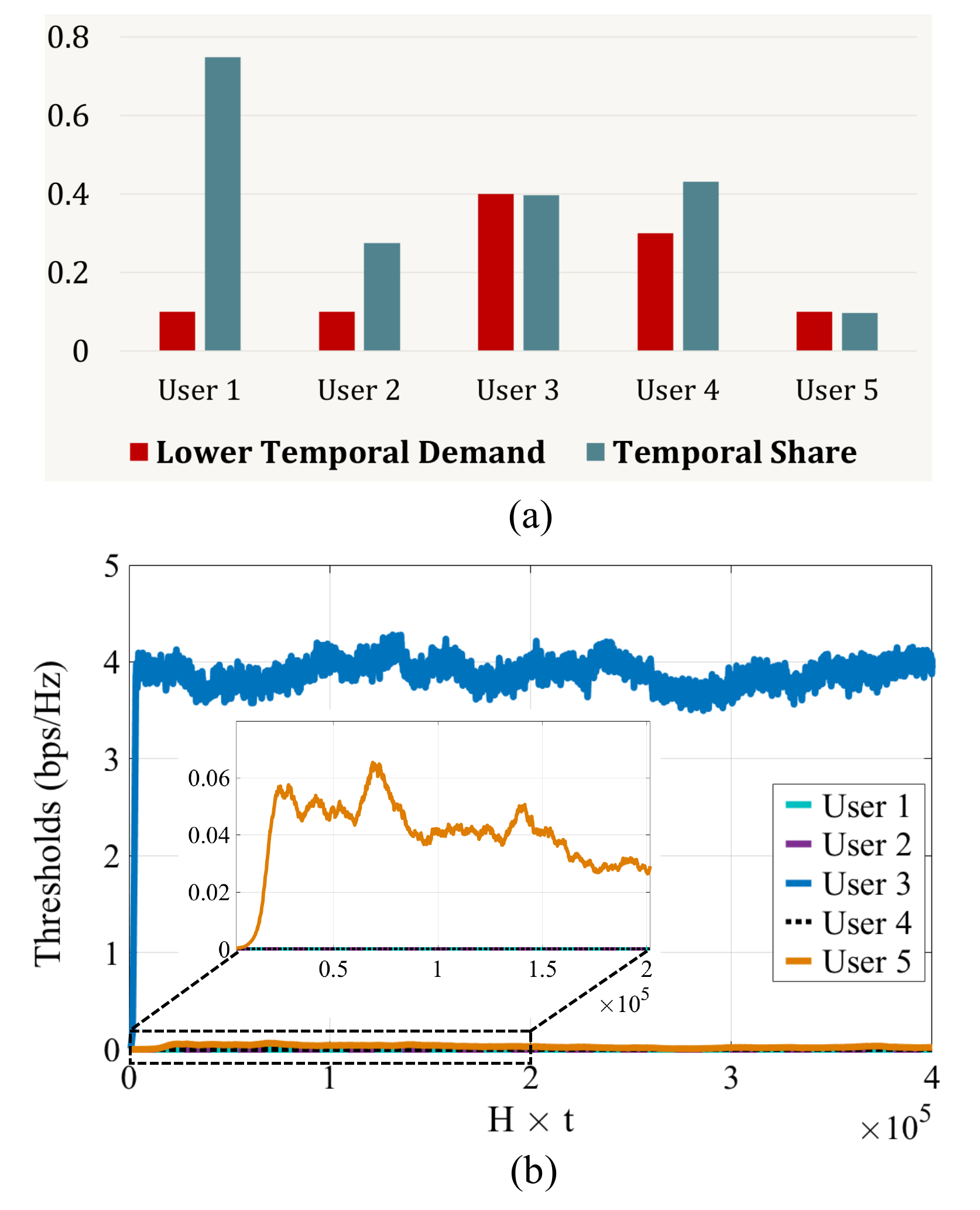}
        \caption{(a) Long-term temporal share of the users in a mobile scenario, (b) The evolution of scheduling thresholds in time. The sampling parameter $H$ is set to $0.1$.}
        \label{fig:temp-share}
        \vspace{-4mm}
\end{figure}
\subsection{Discrete Performance Vectors}
In this section, we consider a NOMA scenario with two users and discrete performance variables and discuss the effectiveness of the method described in Section \ref{sec:discrete}. Due to the small number of virtual users in this scenario, we are able to find the optimal thresholds and  tie-breaking decision analytically. We also use the perturbation method in Section \ref{sec:discrete} to construct a TBS and compare the resulting performance with the optimal utility.  It is shown that the utility from the method proposed in Section \ref{sec:discrete} converges to the optimum utility as $l\to \infty$. Consider a two-user scenario where the performance values are independent discrete random variables distributed as follows:
\begin{align*}
 \resizebox{0.49\textwidth}{!}{
$R_1=
\begin{cases}
0.1~\text{w.p.}~0.5\\
0.2~\text{w.p.}~0.5
\end{cases},~
R_2=
\begin{cases}
0.2~\text{w.p.}~0.5\\
0.3~\text{w.p.}~0.5
\end{cases},~
R_{1,2}=
\begin{cases}
0.1~\text{w.p.}~0.75\\
0.4~\text{w.p.}~0.25
\end{cases}.$}
\end{align*}
Let $\underline{w}_1=0.5$, $\underline{w}_2=0.25$. We first find the optimal thresholds and tie-breaking probability distributions analytically. 
For $\lambda_1=\lambda_2=0$, we have $A_1^Q\leq \frac{7}{16}$. Therefore, $\lambda_1$ must be positive in order to satisfy the temporal demand of user $u_1$. 
Take $\lambda_1$ to $0.1$, Table \ref{tab:discrete} lists all of the possible values for the scheduling measures vector $S^3$ and the choice of the active user. %when using the TBS where $S^3\triangleq \big(S(\mathcal{V}_1,R_1),S(\mathcal{V}_2,R_2),S(\mathcal{V}_{1,2},R_{1,2})\big)$. 
Note that a tie happens when $R^3=(0.1,0.2,0.1)$ and $R^3=(0.2,0.3,0.1)$. In the former, the tie happens among all three virtual users whereas in the latter, the tie is between $\mathcal{V}_1$ and $\mathcal{V}_2$. Let $p^3=(p_1,p_2,p_{1,2})$ denote the tie-breaking distribution. It can be shown that an optimum tie-breaking distribution is $p^3=(\frac{1}{3},\frac{2}{3},0)$ when $R^3=(0.1,0.2,0.1)$ and $p^3=(0,1,0)$ when $R^3=(0.2,0.3,0.1)$ by checking the sufficient conditions in Corollary \ref{cor:CS}. This gives $A_1^Q=0.5$ and $A_2^Q=0.75$ which satisfy the optimality constraints described in Corollary \ref{cor:CS}. As a result, the optimal average system utility is $\frac{45}{160}\approx 0.281$. To evaluate the method proposed in Section \ref{sec:discrete}, we add a vector of independent random variables with distribution $Unif[-\frac{1}{l},\frac{1}{l}]$ to the performance vector and use the perturbed performance values as an input to the TBS. We consider $l \in \{1,2,4,8,16\}$ and use Algorithm \ref{alg:robbins-monro} to obtain the optimal TBS for each value of $l$. Figure \ref{fig:discrete} shows the average system utility for different values of $l$ as well as for the strategy with optimal thresholds and tie-breaking distributions mentioned above. It can be seen that the average system utility converges to the optimal value as $l$ goes to infinity.  
\begin{table}[h]
\centering
\caption{Optimal TBS with discrete performance values}
\scalebox{1}{
\begin{tabular}{lcccc} \toprule
$R^3$ & $(0.1,0.2,0.1)$  & $(0.1,0.2,0.4)$  & $(0.1,0.3,0.1)$  & $(0.1,0.3,0.4)$\\ \midrule
$S^3$ & $(0.2,0.2,0.1)$  & $(0.2,0.2,0.4)$  & $(0.2,0.3,0.1)$  & $(0.2,0.3,0.4)$\\ \midrule
$Q(R^3)$  & $tie$  & $\mathcal{V}_{1,2}$  & $\mathcal{V}_{2}$  & $\mathcal{V}_{1,2}$ \\ \toprule
$R^3$ & $(0.2,0.2,0.1)$  & $(0.2,0.2,0.4)$  & $(0.2,0.3,0.1)$  & $(0.2,0.3,0.4)$ \\ \midrule
$S^3$ & $(0.3,0.2,0.1)$  & $(0.3,0.2,0.4)$  & $(0.3,0.3,0.1)$  & $(0.3,0.3,0.4)$ \\ \midrule
$Q(R^3)$  & $\mathcal{V}_{1}$  & $\mathcal{V}_{1,2}$  & $tie$  & $\mathcal{V}_{1,2}$\\ 
\bottomrule
\end{tabular}
}
\label{tab:discrete}
\end{table}

\begin{figure}[h]
 \centering \includegraphics[width=0.8\linewidth, draft=false]{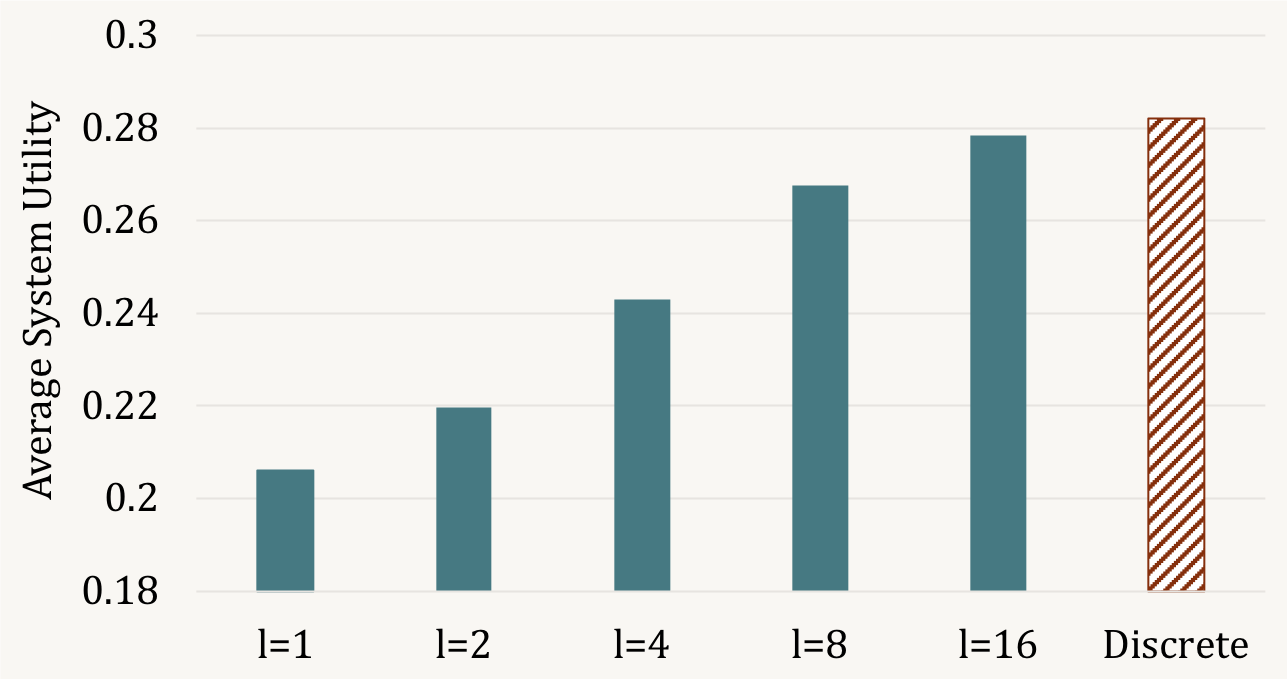}
 \caption{Average system utility for the optimal strategy (tiled red) and the proposed method (filled green). }
 \label{fig:discrete}
\end{figure} 

% \begin{figure}[t]
%  \centering \includegraphics[width=0.8\linewidth, draft=false]{thresholds_discrete_continuous.png}
%  \caption{a}
%  \label{fig:temp-share}
% \end{figure}

\section{Conclusion and Future Work}
In this paper, we have considered scheduling for NOMA systems under temporal demand constraints. We have shown that TBSs achieve optimal system utility and any optimal strategy is equivalent to a TBS. A variable elimination method has been proposed to find the feasible temporal share region for a given NOMA system. We have introduced an iterative algorithm based on the Robbins Monro method which finds the optimal thresholds for the TBS given the user utilities. Lastly, we have provided numerical simulations to validate the proposed approach.

A natural extension to this work is multi-cell scheduling in NOMA systems. The methods proposed here may be extended and applied to centralized and distributed NOMA systems. Particularly, scheduling for multi-cell NOMA systems with limited cooperation is an interesting avenue for future work.

Another direction for future research is NOMA scheduling under short-term fairness constraints. The problem is of interest in delay sensitive applications. Short-term fairness may significantly affect the design of NOMA schedulers. It remains to be seen whether variations of TBSs can achieve near-optimal performance under short-term fairness constraints.

\IEEEpeerreviewmaketitle

\appendix
\subsection{Proof of Lemma \ref{lem:convergence}}

Let $\widetilde{Q}$ be a memoryless and stationary scheduling strategy. Since strategy $\widetilde{Q}$ is memoryless, $Q_t, t\in \mathbb{N}$ is only a function of the realization of performance vector $R^m$ at time $t$. Due to stationarity, the random variables $\mathbbm{1}_{\big\{u_i\in \widetilde{Q}_t(R^{m\times t})\big\}}$ are independent and identically distributed (i.i.d.). From the strong law of large numbers we have: 
\begin{align}
\lim_{t\rightarrow\infty} A_{i,t}^{\widetilde{Q}}~\stackrel{a.s.}{=}~ \mathbb{E}\Big(\mathbbm{1}_{\big\{u_i\in \widetilde{Q}_t(R^{m\times t})\big\}}\Big)\stackrel{(a)}=Pr\Big(u_i\in \widetilde{Q}_t(R^{m})\Big), \label{eq:airtime-stationary-memoryless}
\end{align}
where $(a)$ follows from $\mathbb{E}(\mathbbm{1}_{\mathcal{A}})=Pr(\mathcal{A})$ for any event $\mathcal{A}$. Similarly, the random variables $X_t\triangleq \sum_{j=1}^m R_{j,t}\mathbbm{1}_{\big\{\widetilde{Q}_t(R^{m\times t})=\mathcal{V}_{j}\big\}}$ are i.i.d.. Hence, from the strong law of large numbers we have:
\begin{align}
\lim_{t\rightarrow\infty} U_{\widetilde{Q},t}~\stackrel{a.s.}{=}~ \sum_{j=1}^m\mathbb{E}\Big(R_{j}\mathbbm{1}_{\big\{\widetilde{Q}_t(R^{m})=\mathcal{V}_{j}\big\}}\Big). \label{eq:sysUtility-stationary-memoryless}
\end{align}

% Let $\widetilde{\mathcal{Q}}$ be a stationary and memoryless scheduling policy. Moreover let $A^{i,n}\triangleq\sum_{m=1}^M R^n_{m,i}\mathds{1}_{\{\widetilde{\mathcal{Q}}(\bff{R}^n)=\mathcal{V}_{m}\}}$ and $B^{n}\triangleq\sum_{m=1}^M R^n_{m}\mathds{1}_{\{\widetilde{\mathcal{Q}}(\bff{R}^n)=\mathcal{V}_{m}\}}$. It can be shown that random variables $A^{i,n}$'s are \textit{i.i.d.} over $n$ for every $i$. Similarly, $B^{n}$'s are \textit{i.i.d.} over $n$. By applying law of large numbers to \eqref{eq:user-utility-k} and \eqref{eq:net-utility-k} we have 
% \begin{align}
% U^i_{\widetilde{\mathcal{Q}}}&\triangleq\lim_{k\rightarrow \infty} U^{i,k}_{\widetilde{\mathcal{Q}}}\stackrel{a.s.}{=}\sum_{m=1}^M \mathbb{E}\big[R_{m,i}\mathds{1}_{\{\widetilde{\mathcal{Q}}(\bff{R})=\mathcal{V}_{m}\}}\big], \label{eq:user-utility-stationary-memoryless}\\
% U_{\widetilde{\mathcal{Q}}}&\triangleq\lim_{k\rightarrow \infty} U^k_{\widetilde{\mathcal{Q}}}\stackrel{a.s.}{=}\sum_{m=1}^M \mathbb{E}\big[R_m\mathds{1}_{\{\widetilde{\mathcal{Q}}(\bff{R})=\mathcal{V}_{m}\}}\big].\label{eq:net-utility-stationary-memoryless}
% \end{align}
% Therefore, although there is no guarantee for convergence of sequences $\{U^{i,k}_{\mathcal{Q}}\}_k, \forall i$ and $\{U^{k}_{\mathcal{Q}}\}_k$ when using general policy $\mathcal{Q}$, they would always converge if a stationary and memoryless scheduling policy is used. 

\subsection{Proof of Theorem \ref{th:neq:normal}}

\textbf{Case i) } $w^n=\underline{w}^n=\overline{w}^n, N_{max}=n$

\noindent As an intermediate step, we consider the special case of the scheduling problem when the temporal demand constraints must be satisfied with equality, and all subset of virtual users can be activated, i.e. $\mathsf{V}=2^{\mathcal{U}}$. It turns out that the inclusion of the joint virtual user greatly simplifies the analysis of temporally fair schedulers. 

 First, we prove that if a threshold strategy exists which i) satisfies the temporal constraints, and ii) for which $\lambda_i\in [-2M,2M], \forall i\in [n]$ , then it is optimal, where $M$ is defined in Remark \ref{Rem:bound}. Fix $\epsilon>0$. Let $\epsilon'=2nM\epsilon$. 
Let $\widehat{Q}\in \mathcal{Q}_{TBS}$ be a TBS characterized by the threshold vector $\lambda^n\in [-2M,2M]^n$ and let $Q$ be an arbitrary scheduling strategy. From Equation \eqref{Def:tem_fair} we know that $|A_{i}^{Q}- w_i|\leq \epsilon, \forall i\in [n]$. Also, by assumption, $\lambda_i\leq M, \forall i\in [n]$. As a result, $\lambda_i(A_{i}^{Q}- w_i)+\frac{\epsilon'}{n}\geq 0, \forall i\in [n]$.
We have, \\
\begin{align*}
U_{Q} 
&\leq U_{Q}+\sum_{i=1}^n\big(\lambda_i(A_{i}^{Q}- w_i)\big)+\epsilon' \notag\\
&\leq 
\liminf_{t\rightarrow \infty}
\Bigg[
\frac{1}{t}\sum^t_{k=1}
\sum_{j=1}^m 
\Big(R_{j,k}
\mathbbm{1}_{\big
\{Q_k(R^{m\times k})=\mathcal{V}_{j}\big\}}\Big)\Bigg]+ \notag\\ 
&~~~
\sum_{i=1}^n
\lambda_i\cdot\liminf_{t\rightarrow \infty}
\frac{1}{t}\Bigg[
\sum^t_{k=1}
\Big(
\mathbbm{1}_{\big
\{u_i\in Q_k({R}^{m\times k})\big\}}\Big)\Bigg] -\sum_{i=1}^n\lambda_iw_i+\epsilon' \notag\\
&\stackrel{(a)}{\leq}
\liminf_{t\rightarrow \infty} \frac{1}{t} \sum^t_{k=1} \Bigg[\sum_{j=1}^m 
\Big(R_{j,k}
\mathbbm{1}_{\big
\{Q_k(R^{m\times k})=\mathcal{V}_{j}\big\}}\Big)
\notag
\\
&~~~+\sum_{i=1}^n
\Big(
\lambda_i
\mathbbm{1}_{\big
\{u_i\in Q_k({R}^{m\times k})\big\}}\Big)\Bigg]
-\sum_{i=1}^n\lambda_iw_i+\epsilon'
\notag\\
&=\liminf_{t\rightarrow\infty}
\frac{1}{t}\Bigg[\sum^t_{k=1}\sum_{j=1}^m \Big(\big(R_{j,k}+\sum_{i=1}^n\lambda_i\mathbbm{1}_{\big\{u_i\in\mathcal{V}_{j}\big\}}\big)\mathbbm{1}_{\big\{Q_k(R^{m\times k})=\mathcal{V}_{j}\big\}}\Big)\Bigg]
\notag\\
&~~~-\sum_{i=1}^N\lambda_iw_i+\epsilon' \notag\\
&\stackrel{(b)}{\leq} \liminf_{t\rightarrow\infty}
\frac{1}{t}\Bigg[\sum^t_{k=1}\sum_{j=1}^m \Big(\big(R_{j,k}+\sum_{i=1}^n\lambda_i\mathbbm{1}_{\big\{u_i\in\mathcal{V}_{j}\big\}}\big)\mathbbm{1}_{\big\{\widehat{Q}_k(R^{m\times k})=\mathcal{V}_{j}\big\}}\Big)\Bigg]
\notag\\
&~~~-\sum_{i=1}^n\lambda_iw_i+\epsilon' \notag\\
&\stackrel{(c)}{=}
\liminf_{t\rightarrow \infty}
\Bigg[
\frac{1}{t}\sum^t_{k=1}
\sum_{j=1}^m 
\Big(R_{j,k}
\mathbbm{1}_{\big
\{\widehat{Q}_k(R^{m\times k})=\mathcal{V}_{j}\big\}}\Big)\Bigg] \notag\\ 
&~~~
+\sum_{i=1}^n
\liminf_{t\rightarrow \infty}
\frac{1}{t}\Bigg[
\sum^t_{k=1}
\Big(
\lambda_i
\mathbbm{1}_{\big
\{u_i\in \widehat{Q}_k(R^{m\times k})\big\}}\Big)\Bigg] -\sum_{i=1}^n\lambda_iw_i+\epsilon' \notag\\
&\leq U_{\widehat{Q}}+\overbrace{\sum_{i=1}^n\big(\lambda_i(A_{i}^{\widehat{Q}}- w_i)\big)}^{\leq \epsilon'}+\epsilon'
\\& =U_{\widehat{Q}}+2\epsilon',
\end{align*}
where (a) holds since limit inferior satisfies supper-additivity, (b) holds due to the rearrangement inequality, and finally, (c) follows from the existence of the limit inferior. As a result, we have:
\begin{align*}
U_{Q} \leq U_{\widehat{Q}}+2\epsilon', \forall \epsilon>0, 
\Rightarrow U_Q\leq U_{\widehat{Q}},
\end{align*} 
where equality holds if and only if all of the inequalities above are equalities. Particularly, equality in (b) requires that $Q$ be equivalent with $\widehat{Q}$. 

So far, we have shown that if there exists $\widehat{Q}\in \mathcal{Q}_{TBS}$ is a non-empty set, with $\lambda^n\in [-2M,2M]^n$, then, any optimal strategy is equivalent to a threshold based strategy. In the next step, we show that at least one such $\widehat{Q}$ exists. 

\noindent\textbf{Case i.1) $\sum_{i\in [n]}w_i=1$}

In this case, we show that a threshold based strategy with $\lambda_i\in [-2M,-M], \forall i\in [n]$ exists.
Note that if $\lambda_i\leq -M, \forall i\in [n]$, then only the individual users (i.e. $\mathcal{V}_i=\{u_i\}, i\in [n]$) will be chosen by the threshold strategy. The reason is that the scheduling measures for the individual users are larger than that of joint users 
with probability one due to Remark \ref{Rem:bound}.
Furthermore, from \cite{liu-infocom}, it is known that when individual users are chosen, one can find a set of thresholds such that $A_i^Q=w_i$ with probability one for $\sum_{i\in [n]}w_i=1$. This shows the existence of suitable thresholds in Case i.1.

\noindent\textbf{Case i.2) $\sum_{i\in [n]}w_i>1$}

To prove existence in this case, we use the following n-dimensional extension of the mean value theorem.

\begin{Lemma}[\textbf{Poincar\'{e}-Miranda}]
\label{thm:poincare}
Let $n\in \mathbb{N}$. Consider the set of continuous functions $f_i: \mathbb{R}^n \rightarrow \mathbb{R}, i \in [n]$. Assume that for each function $f_i, i\in[n]$, there exists positive reals $M^+_i$ and $M^-_i$, such that $f_i(x^n)>0$ if $x_i= M^+_i$ and $f_i(x^n)<0$ if $x_i= M^-_i$. Then, the function $f^n=(f_1,f_2,\cdots,f_n)$ has a root in the 
 n-dimensional cube $\prod_{i=1}^n[-M^-_i,M^+_i]$. Alternatively:
\begin{align*}
 \exists x^*_1,\ldots,x^*_n\in  \prod_{i=1}^n[-M^-_i,M^+_i] : f_i(x^*_1,\ldots,x^*_n)=0,\forall i\in[n]. 
\end{align*}
\label{Lem:PM}
\end{Lemma}
We provide the proof when $0<w_i<1, i\in [n]$. Take $f_i(\lambda^n)\triangleq A_i^{Q_{TBS}}-w_i, \forall i\in [n]$. Then, $f_i$ are continuous functions of $\lambda^n$. Next, we find a set of thresholds $(M_i^+, M_i^-), i\in [n]$ satisfying the conditions of Lemma \ref{Lem:PM}. Note that if $\lambda_i=M$, $u_i\in Q_{TBS}(R^n)$ with probability one.  To see this, let $\mathcal{V}_j$ be a virtual user such that $u_i\notin \mathcal{V}_j$ and let $\mathcal{V}'_j= \mathcal{V}_j\cup \{u_i\}$. Then, 
\begin{align*}
&P(S(\mathcal{V}_j,R_{t,j})\leq S(\mathcal{V}'_j,R_{t,j}))
\\&= P(R_{t,j}+\sum_{i=1}^n \lambda_i \mathbbm{1}_{\{u_i\in\mathcal{V}_{j}\}}\leq R_{t,j'}+\sum_{i=1}^n \lambda_i \mathbbm{1}_{\{u_i\in\mathcal{V}_{j}\}}+M)
\\& = P(R_{t,j}- R_{t,j'}\leq M)=1,
\end{align*}
where the last equality follow from Remark \ref{Rem:bound}. As a result, $A_i^{Q_{TBS}}-w_i=1-w_i>0$. Hence, $M_i^+=M$ satisfies the conditions of Lemma \ref{Lem:PM}.
Next, we construct $M_i^-, i\in [n]$. Note that by assumption $e\triangleq \frac{\sum_{i\in [n]}w_i-1}{n}>0$. Furthermore, it is straightforward to show that there exists $\alpha^n>0$ such that $\sum_{i\in [n]}\alpha_i=1$ and $w_i-\alpha_i e>0, i\in [n]$. Define $w'_i=w_i-\alpha_i e,i\in [n]$. Then, by construction,  $\sum_{i\in [n]} w'_i=1$. By similar arguments as in the case i, for any fixed $i\in [n]$, there exists $\lambda_i\in [-2M,-M]$ such that $A_i^Q=w'_i<w_i$. So, $A_i^Q-w'_i<0$. Consequently, $M_i^-=\lambda_i$ satisfies the condition that $f_i(\lambda^n)<0, \lambda_i=M_i^-, \forall i\in [n]$. By Lemma \ref{Lem:PM}, there exists $\lambda^n$ such that $A_i^Q=w_i, i\in [n]$ simultaneously.

\noindent\textbf{Case ii) $w^n=\underline{w}^n=\overline{w}^n, N_{max}<n$}

Similar to the proof for C-NOMA systems, it can be shown that if a TBS exists, then it is optimal, and that any optimal strategy is equivalent to the TBS. The proof of existence of a TBS in the case that $\sum_{i\in [n]}w_i=1$ is similar to the proof of Case i.1. However, the proof requires additional arguments when $\sum_{i\in [n]}w_i>1$. 
 We use the  following extension of Lemma \ref{Lem:PM}.  
\begin{Lemma}[\textbf{Avoiding Cones Conditions \cite{cones}}]

Let $n\in \mathbb{N}$. Consider the set of continuous functions $f_i: \mathbb{R}^n \rightarrow \mathbb{R}, i \in [n]$. Assume that for each function $f_i, i\in[n]$, there are positive reals $M^+_i$ and $M^-_i$ such that i) for any point $x^n$ such that  $x_i=M^+_i$, either the function $f_i$ is positive or $\exists j\neq i: f_j(x^n)\neq 0$, and ii) For any point $x^n$ such that $x_i=-M^-_i$, either the function $f_i$ is negative or $\exists j\neq i: f_j(x^n)\neq 0$. Then, the function $f^n=(f_1,f_2,\cdots,f_n)$ has a root in the 
 n-dimensional cube $\prod_{i=1}^n[-M^-_i,M^+_i]$. Alternatively:
\begin{align*}
 \exists x^*_1,\ldots,x^*_n\in  \prod_{i=1}^n[-M^-_i,M^+_i] : f_i(x^*_1,\ldots,x^*_n)=0,\forall i\in[n]. 
\end{align*}
\label{Lem:AE}
\end{Lemma}
Take $f_i(\lambda^n)\triangleq A_i^{Q_{TBS}}-w_i, \forall i\in [n]$. The set of thresholds $M^-_i=-\lambda_i^*$ can be constructed by the precise same method used in Case i.2. Let $l_{max}$ be the total maximum number of users in the joint users of the B-NOMA system. We provide the proof for $l_{max}=2$ and $1<\sum_{i\in [n]}w_i < l_{max}$. The general proof follows by similar arguments.
In this case, for a fixed $i$, let $\lambda_i=2M$. We claim that $f_i(\lambda^n)=A_i^Q- w_i>0, \forall \lambda^n:\lambda_i=2M$. Define the following partition of the threshold space $(\Lambda_1,\Lambda_1^c)$, where:
\begin{align*}
\Lambda_1= \left\{\lambda^n| \lambda_i=1 \text{ \& } \lambda_j<M, j\neq i\right\}.
\end{align*}
If $\lambda^n\in \Lambda_1$, then, $u_i\in Q_{TBS}(R^n)$ with probability one. To see this, let $\mathcal{V}_j$ be a virtual user that $u_i\notin \mathcal{V}_j$ and let $u_j \in \mathcal{V}_j$. Define $\mathcal{V}'_j=\mathcal{V}_j\cup \{u_i\}-\{u_j\}$, then 
\begin{align*}
&P(S(\mathcal{V}_j,R_{t,j})\leq S(\mathcal{V}'_j,R_{t,j}))
\\&= P(R_{t,j}+\sum_{i=1}^n \lambda_i \mathbbm{1}_{\{u_i\in\mathcal{V}_{j}\}}\leq R_{t,j'}+\sum_{i=1}^n \lambda_i \mathbbm{1}_{\{u_i\in\mathcal{V}_{j}\}}+2M)
\\& = P(R_{t,j}- R_{t,j'}+\lambda_j\leq 2M)=1.
\end{align*}
So, $f_i(\lambda^n)=1>w_i$.
On the other hand, if $\lambda^n \in \Lambda^c_1$, then it is straightforward to show that only joint virtual users with $l_{max}$ users are chosen. As a result, $\sum_{i\in [n]}A_i^Q=l_{max}$. So, $\sum_{i\in [n]}f_i(\lambda^n)=\sum_{i\in [n]}(A_i^Q-w_i)= l_{max}-\sum_{i\in [n]} w_i >0$. Consequently, there exists $j\in [n]$ such that $f_j\neq 0$.

%We provide the proof when $w_i-\frac{\sum_{j\in [n]}w_j-1}{n}>0, i\in [n]$. The general proof follows by similar arguments. 
%Let $f_i(\lambda^n)={A}^i_{Q_{TBS}(\lambda^n)}- w_i, i\in \mathbb{N}$. We show that $f=(f_i)_{i\in [n]}$ satisfies the conditions in Lemma \ref{Lem:PM}. First, we find suitable values for $M_i^+, i\in [n]$. Fix $i\in \mathbb{N},$ note that if $\lambda_i= M$, where $M$ is defined in Remark \ref{Rem:bound}, then $w_i=1$ regardless of the value of $\lambda_j, j\in [n]-\{i\}$. As a result, we set $M_i^+=M, i\in [n]$. Next, we find suitable values for $M_i^-, i\in [n]$. Let $e= \frac{\sum_{w_i-1}}{n}$. Note that $e>0$ by the assumption that $\sum_{i\in [n]}w_i>1$. 
 % Let $M_i^-= \lambda^*_i, i\in [n]$. Then, it is straightforward to check that the conditions in Lemma \ref{Lem:PM} are satisfied and there exist thresholds such that $f_i(\lambda^n)=0$ simultaneously. Alternatively,  
\noindent\textbf{Case iii)} $\underline{w}^n\neq \overline{w}^n, N_{max}\leq n$

We provide a sketch of the proof. 
First, we argue that an optimal strategy exists. Let $\mathcal{W}=\{w^n: \underline{w}_i \leq w_i\leq \overline{w}_i, \mathcal{Q}\neq \phi\}$ be the set of all feasible weights which satisfy the temporal constraints. Let $U^*(w^n), w^n\in \mathcal{W}$ be the optimal utility achieved by any arbitrary scheduling strategy with weight vector $w^n$ under equality constraints. The function $U^*(\cdot)$ is a continuous function. Furthermore, the subspace $\mathcal{W}$ is a bounded convex polytope in $\mathbb{R}^n$. As a result, 
\begin{align*}
{w}^{*,n}=argmax_{{w}^n\in {\mathcal{W}}} U^*(w^n)
\end{align*}
exists. On the other hand, from the first part of the proof (the $\underline{w}^n=\overline{w}^n$ case), we know that the optimal strategy under equality temporal constraints with weight vector $w^{* n}$ is equivalent to a TBS. Consequently, the optimal strategy exists and is equivalent to a TBS strategy. 

\subsection{Proof of Theorem \ref{th:converge}}
Condition (4) follows from the properties of the Harmonic series. We provide the proof for condition (2). We showed in Theorem \ref{th:neq:normal} that the optimal threshold vector ${\lambda^*}^n$  exist. Let $\epsilon^n$ be an arbitrary vector of real numbers. Define $b^*_i$ and $b_i, i\in [n]$ as the resulting temporal shares for the optimal TBS with threshold vector ${\lambda^*}^n$ and the temporal shares for $Q_{TBS}(\lambda^n)$, respectively, where $\lambda^n={\lambda^*}^n+\epsilon^n$.
Let $\mathcal{A}^*_{j}$ and $\mathcal{A}_{j}, j\in [m]$ be the event that virtual user $j$ is activated by $Q_{TBS}({\lambda^*}^n)$ and $Q_{TBS}(\lambda^n)$, respectively. Then, we need to show that:
\begin{align}
(\epsilon^n)^T\left({b^*}^n-b^n\right)>0,
\label{eq:suff1}
\end{align}
where $b^*_i= \sum_{j:u_i\in \mathcal{V}_j}P(\mathcal{A}^*_{j})$, and $b_i= \sum_{j:u_i\in \mathcal{V}_j}P(\mathcal{A}_{j})$. Equation \eqref{eq:suff1} can be written as:
\begin{align}
\sum_{i\in [n]}\epsilon_i(b^*_i-b_i)>0.
\label{eq:suff2}
\end{align}
Note that by the law of total probability
\begin{align*}
& b^*_i= \sum_{k\in [m]}\sum_{j:u_i\in \mathcal{V}_j}P(\mathcal{A}^*_{j}\bigcap \mathcal{A}_{k}),\\
& b_i= \sum_{k\in [m]}\sum_{j:u_i\in \mathcal{V}_j}P(\mathcal{A}_{j}\bigcap \mathcal{A}^*_{k}).
\end{align*}
As a result, we need to show that
\begin{align*}
&\sum_{i\in [n]}\epsilon_i\left(\sum_{k\in [m]}\sum_{j:u_i\in \mathcal{V}_j}P\left(\mathcal{A}^*_{j}\bigcap \mathcal{A}_{k}\right)
- \sum_{k\in [m]}\sum_{j:u_i\in \mathcal{V}_j}P\left(\mathcal{A}_{j}\bigcap \mathcal{A}^*_{k}\right)\right)>0\\
& \Leftrightarrow 
\sum_{k\in [m]}\sum_{i\in [n]}\sum_{j:u_i\in \mathcal{V}_j}\epsilon_i\left(P\left(\mathcal{A}^*_{j}\bigcap \mathcal{A}_{k}\right)
- P\left(\mathcal{A}_{j}\bigcap \mathcal{A}^*_{k}\right)\right)>0\\
& \Leftrightarrow 
\sum_{k\in [m]}
\sum_{j\in [m]}
\sum_{i: u_i\in \mathcal{V}_j}
\epsilon_i\left(P\left(\mathcal{A}^{*}_{j}\bigcap \mathcal{A}_{k}\right)
- P\left(\mathcal{A}_{j}\bigcap \mathcal{A}^*_{k}\right)\right)>0\\
&\Leftrightarrow 
\sum_{k\in [m]}
\sum_{j\in [m]}
\left(\sum_{i: u_i\in \mathcal{V}_j}
\epsilon_i\left(P\left(\mathcal{A}^{*}_{j}\bigcap \mathcal{A}_{k}\right)
- P\left(\mathcal{A}_{j}\bigcap \mathcal{A}^*_{k}\right)\right)\right)>0\\
& \Leftrightarrow 
\sum_{k\in [m]}
\sum_{j\in [m]}
\sum_{i: u_i\in \mathcal{V}_j}
\epsilon_iP\left(\mathcal{A}^{*}_{j}\bigcap \mathcal{A}_{k}\right)
\\&
\qquad \qquad \qquad \quad- 
\sum_{k\in [m]}
\sum_{j\in [m]}
\sum_{i: u_i\in \mathcal{V}_k} 
\epsilon_i P\left(\mathcal{A}^*_{j}\bigcap \mathcal{A}_{k}\right)
>0\\
& \Leftrightarrow 
\sum_{k\in [m]}
\sum_{j\in [m]}
P\left(\mathcal{A}^{*}_{j}\bigcap \mathcal{A}_{k}\right)
\left(\sum_{i: u_i\in \mathcal{V}_j}
\epsilon_i\right)
\\&
\qquad \qquad \qquad \quad- 
\sum_{k\in [m]}
\sum_{j\in [m]}
 P\left(\mathcal{A}^*_{j}\bigcap \mathcal{A}_{k}\right)
\left(\sum_{i: u_i\in \mathcal{V}_k} 
\epsilon_i\right)
>0.
\end{align*}
Let $e_{j}=\sum_{i: u_i\in \mathcal{V}_j}\epsilon_i, j\in [m]$. Note that $e_j$ is the perturbation of the scheduling measure of the virtual user $j$ defined in Definition \ref{def:U_TBS} resulting from changing ${\lambda^*}^n$ to $\lambda^n$. In fact the scheduling measure can be written as 
$S\big(\mathcal{V}_j,R_{j}\big)=R_j+\sum_{i:u_i\in\mathcal{V}_j }\lambda^*_i+ e_j$.
We need to show that:
\begin{align*}
&  
\sum_{k\in [m]}
\sum_{j\in [m]}
e_j P\left(\mathcal{A}^{*}_{j}\bigcap \mathcal{A}_{k}\right)- 
\sum_{k\in [m]}
\sum_{j\in [m]}
e_k P\left(\mathcal{A}^*_{j}\bigcap \mathcal{A}_{k}\right)
>0\\
& \Leftrightarrow 
\sum_{k\in [m]}
\sum_{j\in [m]}
(e_j-e_k) \left(P\left(\mathcal{A}^*_{j}\bigcap \mathcal{A}_{k}\right)
-
P\left(\mathcal{A}^*_{k}\bigcap \mathcal{A}_{j}\right)\right)
>0
\end{align*}
We claim that $e_j-e_k$ and $P(\mathcal{A}^*_{j}\bigcap \mathcal{A}_{k})
-
P(\mathcal{A}^*_{k}\bigcap \mathcal{A}_{j})$ have the same sign for all $j,k\in [m]$. To see this, note that if $e_j>e_k$ then the threshold for virtual user $j$ is increased more than that of virtual user $k$ after perturbing ${\lambda^*}^n$ by $\epsilon^n$. As a result, it can be shown that $P(\mathcal{A}^*_{j}\bigcap \mathcal{A}_{k})
>
P(\mathcal{A}^*_{k}\bigcap \mathcal{A}_{j})$. Roughly speaking, this can be interpreted as follows: if the threshold   for virtual user $j$ is increased more than that of virtual user $k$, then its temporal share increases more than that of $k$ as well. A similar argument can be provided for $e_j<e_k$. This completes the proof.

\subsection{Proof of Lemma \ref{lem:concave}}
Fix the pair of vectors of temporal demands $(\underline{w}^n.\overline{w}^n)$ and  $({\underline{w}'}^n,{\overline{w}'}^n)$ and $\alpha\in [0,1]$. We need to show the following inequality:
\begin{align}
U^*_{{\underline{w}''}^n, {\overline{w}''}^n}\geq 
\alpha U^*_{{\underline{w}}^n, {\overline{w}}^n}
+(1-\alpha)U^*_{{\underline{w}'}^n, {\overline{w}'}^n},
\label{eq:conc}
\end{align}
where ${\underline{w}''}^n=\alpha {\underline{w}}^n+(1-\alpha){\underline{w}'}^n$ and 
${\overline{w}''}^n= \alpha {\overline{w}}^n+ (1-\alpha)  {\overline{w}'}^n $. Let $Q^{TBS}$ be the optimal strategy for the temporal constraints $(\underline{w}^n.\overline{w}^n)$. Also, let ${Q'}^{TBS}$ be the optimal strategy for the temporal constraints $({\underline{w}'}^n,{\overline{w}'}^n)$ . Define the strategy $Q''$ as follows: for $\alpha$ fraction of the time-slots, the strategy chooses the active user based on $Q^{TBS}$ and for $(1-\alpha)$ fraction of the time it uses ${Q'}^{TBS}$ to choose the active user. It is straightforward to verify that the resulting temporal shares are between $({\underline{w}''}^n,{\overline{w}''}^n)$. Furthermore, the resulting utility from $Q''$ is $U''=\alpha U^*_{{\underline{w}}^n, {\overline{w}}^n}
+(1-\alpha)U^*_{{\underline{w}'}^n, {\overline{w}'}^n}$. 
By definition we have $U''\leq U^*_{{\underline{w}''}^n, {\overline{w}''}^n}$. This completes the proof.

\subsection{Proof of Theorem \ref{th:discrete}}
We provide a sketch of the proof. 
%Note that by definition we have $U_{\epsilon} \leq U^*, \epsilon \in \mathbb{R}$. We will show that for any $\delta>0$, there exists $l \in \mathbb{N}$ such that for any $k\geq l$,  $U^*\leq U_{\frac{1}{k}}+\delta$. 
Fix $l\in \mathbb{N}$. We construct a genie-assisted strategy $\widetilde{Q}_{\frac{1}{l}}$ for the setup $\Omega_{\frac{1}{l}}$. The strategy uses the output of $Q^*$ as side-information assuming that the output is provided using a genie. At time-slot $t$ the strategy $\widetilde{Q}_{\frac{1}{l}}$ for $\Omega_{\frac{1}{l}}$ activates the same virtual user $\mathcal{V}_j$ which the strategy $Q^*$ activates for $\Omega_{0}$. Let $\widetilde{U}_{\frac{1}{l}}$ be the average utility of $\widetilde{Q}_{\frac{1}{l}}$. Then, 
$U^*\leq \widetilde{U}_{\frac{1}{l}}+\frac{1}{{l}}$. This is true since the utility in $\Omega_0$ and $\Omega_{\frac{1}{l}}$ differ in a $Unif[-\frac{1}{l},\frac{1}{l}]$ variable. Using the rearrangement inequality as in the proof of Theorem \ref{th:neq:normal}, it can be shown that $P(\widetilde{U}_{\frac{1}{l}}\leq U^*_{\frac{1}{l}})=1$, where  $U^*_{\frac{1}{l}}$ is the utility from applying $Q_{\frac{1}{l}}$ to $\Omega_{\frac{1}{l}}$ and $Q_{\frac{1}{l}}$ is defined in the theorem statement. Furthermore, let ${U}_{\frac{1}{l}}$ be the utility from ${Q}_{\frac{1}{l}}$ applied to $\Omega_0$ . Then, $U^*_{\frac{1}{l}}\leq {U}_{\frac{1}{l}}+\frac{1}{l}$ by similar arguments. As a result, 
\begin{align*}
&P(U^*\leq U_{\frac{1}{l}}+\frac{2}{l})\geq P(U^*\leq U^*_{\frac{1}{l}}+\frac{1}{l})
\geq P(U^*\leq \widetilde{U}_{\frac{1}{l}}+\frac{1}{l})
=1\\
& \Rightarrow P(U^*\leq U_{\frac{1}{l}}+\frac{2}{l})=1.
\end{align*}
On the other hand $P(U^*\geq U_{\frac{1}{l}})=1$ by definition of $U^*$. The proof is completed by taking $l$ to infinity.

\bibliographystyle{IEEEtran}
\bibliography{reference}
\end{document}